\documentclass[prb,aps,amssymb,twocolumn,superscriptaddress]{revtex4-1}
\usepackage{amsmath}
\usepackage{amssymb}
\usepackage{amsthm}
\usepackage{amsfonts}
\usepackage{listings}
\usepackage{braket}
\usepackage{enumerate}
\usepackage{latexsym}
\usepackage[dvips]{graphicx}
\usepackage{threeparttable}
\usepackage{multirow}
\usepackage{dsfont}
\usepackage{enumerate}
\usepackage{placeins}
\usepackage{tabularx}
\usepackage{hyperref}
\usepackage{psfrag}
\usepackage{comment}

\usepackage{bm}
\usepackage{color}

\newcommand{\beq}{\begin{equation}}
\newcommand{\eneq}{\end{equation}}
\newcommand{\bs}[1]{\boldsymbol{#1}}

\begin{document}

\title{Theory of inversion-$\mathbb{Z}_{4}$ protected topological chiral hinge states and its applications to layered antiferromagnets
}

\author{Yutaro Tanaka}
\affiliation{
 Department of Physics, Tokyo Institute of Technology, 2-12-1 Ookayama, Meguro-ku, Tokyo 152-8551, Japan
}
\author{Ryo Takahashi}
\affiliation{
 Department of Physics, Tokyo Institute of Technology, 2-12-1 Ookayama, Meguro-ku, Tokyo 152-8551, Japan
}
\author{Tiantian Zhang}
\affiliation{
 Department of Physics, Tokyo Institute of Technology, 2-12-1 Ookayama, Meguro-ku, Tokyo 152-8551, Japan
}
\affiliation{
 TIES, Tokyo Institute of Technology, 2-12-1 Ookayama, Meguro-ku, Tokyo 152-8551, Japan
}
\author{Shuichi Murakami}
\affiliation{
 Department of Physics, Tokyo Institute of Technology, 2-12-1 Ookayama, Meguro-ku, Tokyo 152-8551, Japan
}
\affiliation{
 TIES, Tokyo Institute of Technology, 2-12-1 Ookayama, Meguro-ku, Tokyo 152-8551, Japan
}

\begin{abstract}
We study positions of chiral hinge states in higher-order topological insulators (HOTIs) with inversion symmetry.
First, we exhaust all possible configurations of the hinge states in the HOTIs in all type-I magnetic space groups with inversion symmetry by studying dependence of the sign of the surface Dirac mass on surface orientations. In particular, in the presence of glide symmetry, for particular surface orientations, the surface Dirac mass changes sign by changing the surface terminations. By applying this result to a layered antiferromagnet (AFM), we find a difference in the hinge states between the cases with an even and odd number of layers. In the case of an even number of layers, which does not preserve inversion symmetry, positions of hinge states are not inversion symmetric.
Nonetheless, these inversion-asymmetric hinge states result from the bulk topology. We show that  their inversion-asymmetric configurations are uniquely determined from the symmetries and the topological invariant.
\end{abstract}

\maketitle

\section{introduction}
The discovery of topological insulators (TIs) have triggered intensive studies on topological aspects in electronic structures of solids \cite{RevModPhys.82.3045, RevModPhys.83.1057}.
Three-dimensional (3D) TIs have gapless surface states which are protected by time-reversal ($\mathcal{T}$) symmetry.
On the other hand, these surface states can be gapped by breaking $\mathcal{T}$ symmetry while preserving topological properties of the bulk; such TIs are called magnetic TIs. 
Among magnetic TIs, those characterized by the quantized value of the Chern-Simons axion angle $\theta=\pi$  \cite{PhysRevLett.58.1799, PhysRevB.78.195424, PhysRevLett.102.146805, PhysRevB.81.205104, PhysRevB.89.165132, PhysRevB.93.045115, PhysRevB.98.245117, wieder2018axioninsulatorpump, PhysRevLett.122.256402, varnava2019axion, wieder2020dynamical} are known as axion insulators (AXIs). The simplest AXIs have been proposed in inversion ($\mathcal{I}$) symmetric TIs in an external magnetic field and in a TI doped with magnetic atoms without breaking $\mathcal{I}$ symmetry. $\mathcal{I}$ symmetry pins the axion angle $\theta$ to $\pi$ even when $\mathcal{T}$ symmetry is broken.  AXIs have the quantized magnetoelectric effect from the nontrivial axion angle $\theta=\pi$  \cite{PhysRevB.78.195424, PhysRevLett.102.146805, PhysRevB.81.205104}, and the surfaces of AXIs have a half quantum Hall effect \cite{PhysRevB.78.195424, PhysRevB.98.245117}. 
Recently,  EuIn$_{2}$As$_{2}$ \cite{PhysRevLett.122.256402} and ${\rm Mn}{\rm Bi}_{2} {\rm Te}_{4}$ \cite{PhysRevLett.122.206401} have been proposed as layered antiferromagnetic  TIs, and then they have been studied extensively \cite{ otrokov2019prediction, gong2019experimental, li2019intrinsic, PhysRevB.100.121104, hu2020van, wu2019natural, PhysRevX.9.041039, deng2020quantum,regmi2019temperature,zhang2019plane, PhysRevLett.124.136407, xu2020high}. In ${\rm Mn}{\rm Bi}_{2} {\rm Te}_{4}$, the combined symmetry $\mathcal{S}= \mathcal{T} \tau_{1/2}$ leads to a $\mathbb{Z}_2$ topological classification in the absence of $\mathcal{T}$ symmetry \cite{PhysRevB.81.245209}, where $\mathcal{T}$ and $\tau_{1/2}$ represent a $\mathcal{T}$ operator and a half-translation one respectively.

Higher-order TIs (HOTIs) have been proposed as a new class of TIs  \cite{PhysRevLett.108.126807,PhysRevLett.110.046404,PhysRevB.92.085126, benalcazar2017quantized, PhysRevB.96.245115,PhysRevLett.119.246402, fang2017rotation, PhysRevB.97.205135, PhysRevB.97.241405, PhysRevB.98.081110, PhysRevB.98.201114, PhysRevB.98.205129, PhysRevB.98.245102, schindler2018higher,  PhysRevLett.119.246401, PhysRevB.97.205136, PhysRevB.98.245102,schindler2018higherbismuth, wang2018higher, ezawa2019scientificrepo,PhysRevB.99.041301,PhysRevX.9.011012,okugawa2019second,yue2019symmetry,PhysRevLett.122.076801,PhysRevB.98.205147,PhysRevLett.121.196801,PhysRevB.98.235102,PhysRevLett.123.016805,PhysRevLett.123.073601,serra2018observationnature555,peterson2018quantizedNature7695,imhof2018topolectricalnatphys,PhysRevB.99.195431,PhysRevLett.123.016806,sheng2019two,agarwala2019higher,PhysRevLett.123.036802,chen2019higher,PhysRevB.98.035147,PhysRevResearch.2.013083, ghosh2019engineering, hirayama2020higher, ghosh2019engineering, chen2020universal}. 3D HOTIs have topological one-dimensional states along hinges of the systems, which are called hinge states. 
Hinge states arise from nontrivial higher-order topology of the bulk. Among various classes of HOTIs, one class of HOTIs is protected by $\mathcal{I}$ symmetry \cite{wang2018higher,PhysRevLett.122.256402, PhysRevB.98.205129,PhysRevB.97.205136,PhysRevB.98.245102,schindler2018higherbismuth,wieder2018axioninsulatorpump,ezawa2019scientificrepo}. The HOTIs without $\mathcal{T}$ symmetry have chiral hinge states (CHSs) \cite{PhysRevLett.122.256402, PhysRevB.98.205129,PhysRevB.97.205136,PhysRevB.98.245102,wieder2018axioninsulatorpump}, whereas the HOTIs with $\mathcal{T}$ symmetry have helical hinge states \cite{ezawa2019scientificrepo,schindler2018higherbismuth,wang2018higher, zhou2020glide}. In the magnetic systems only with $\mathcal{I}$ symmetry, the topological phases are characterized by three weak $\mathbb{Z}_{2}$ indices  and the strong $\mathbb{Z}_{4}$ index $\mu_{1}$ \cite{PhysRevX.7.041069, po2017symmetry,PhysRevX.8.031070, PhysRevB.98.115150,  tang2019comprehensive,tang2019efficient}, and CHSs always appear when all the weak indices vanish and $\mu_{1}=2$ \cite{PhysRevB.101.115120, PhysRevResearch.2.013300}. 
Therefore, magnetic TIs with $\mathcal{I}$ symmetry and $\mu_{1}=2$ have CHSs, and  
remarkably, they are AXIs with $\theta=\pi$ \cite{wieder2018axioninsulatorpump, PhysRevLett.122.256402, PhysRevB.98.245117, wieder2020dynamical}. 
The positions of the CHSs found are always $\mathcal{I}$ symmetric \cite{PhysRevLett.122.256402, PhysRevB.98.205129,PhysRevB.97.205136,PhysRevB.98.245102,wieder2018axioninsulatorpump}.

In this paper, we find all the possible patterns of positions of CHSs in HOTIs from $\mathcal{I}$ symmetry. We exhaust all the patterns by studying dependence of the sign of the surface Dirac mass on surface orientations for each space group. In particular, 
in the presence of glide symmetry, for particular surface orientations, the surface Dirac mass changes sign by changing the surface terminations. By applying this result to a layered antiferromagnet (AFM), we find a difference in the hinge states between the cases with an even and odd number of layers, and we find emergence of $\mathcal{I}$-asymmetric hinge states (IAHS) in the case of an even number of layers. Moreover, we show that IAHS result from the bulk $\mathbb{Z}_{4}$ topology protected by $\mathcal{I}$ symmetry, and they generally appear in antiferromagnets (AFMs) with an even number of layers and the non-trivial $\mathbb{Z}_{4}$ index. In addition, we show that their $\mathcal{I}$-asymmetric positions are uniquely determined from an interplay of the symmetries and topology, which is also discussed in the axion insulator EuIn$_{2}$As$_{2}$ \cite{PhysRevLett.122.256402,regmi2019temperature,zhang2019plane}.

The organization of the paper is as follows. In Sec.~\ref{sec:typeImagneticspace}, we find all the possible patterns of positions of CHSs in HOTIs in all type-I MSGs with $\mathcal{I}$ symmetry. In Sec.~\ref{sec:hingestatesinlayered}, we discuss CHSs in layered antiferromagnetic HOTIs, and then we show that the positions of the CHSs depend on the parity of the number of layers $N$. In Sec.~\ref{sec:modelcalculations}, we perform calculations on the tight-binding model to confirm our theory in Sec.~\ref{sec:hingestatesinlayered}. Our conclusion is given in Sec.~\ref{sec:conclusion}.

\section{Type-I Magnetic space groups with inversion symmetry}\label{sec:typeImagneticspace}\label{sec2:chiral_antiferro}
\begin{figure}
\includegraphics[width=1\columnwidth]{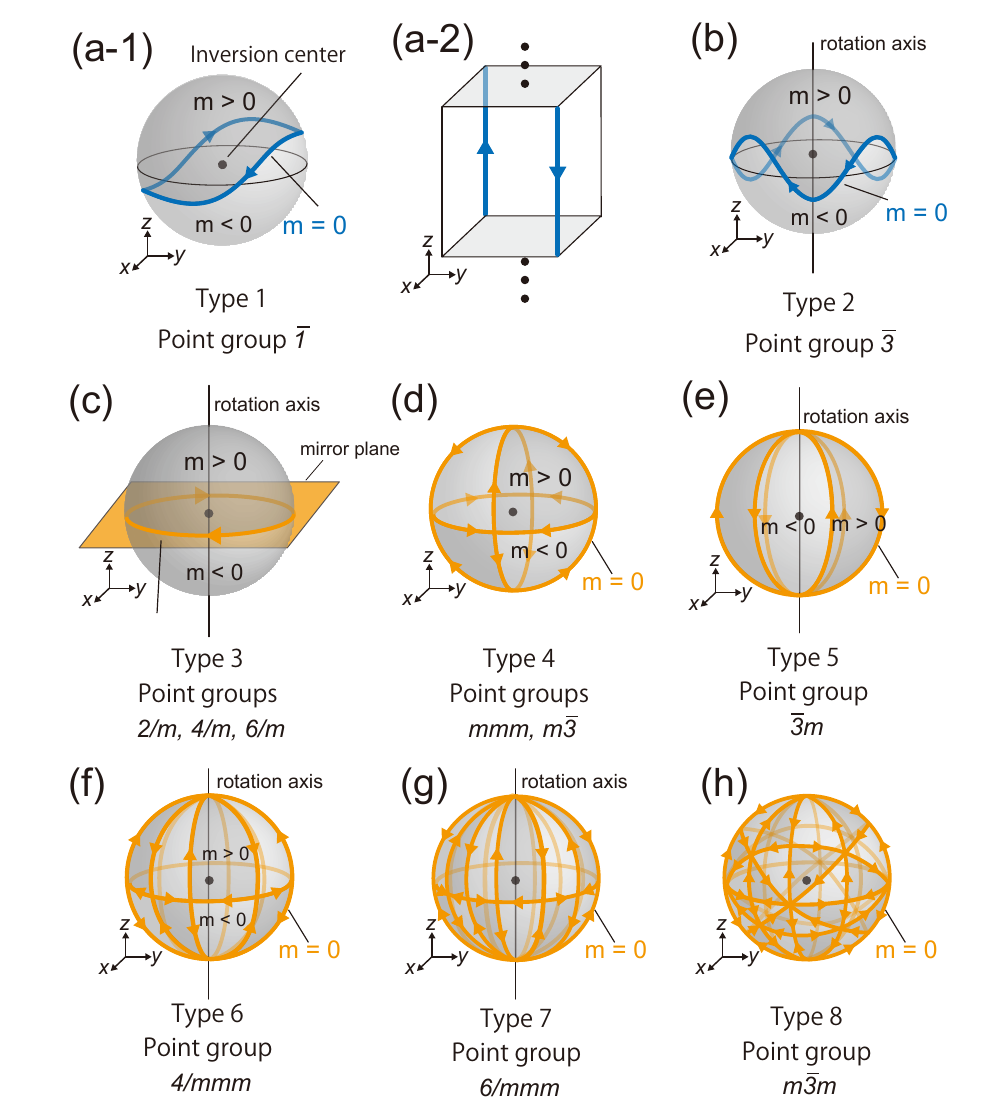}
\caption{Eight types of the distributions of the mass term and CHSs as a function of the surface normal $\bs{n}$ shown on a sphere with $\mathcal{I}$ symmetry. (a-1) Type 1:~The CHS forms a loop protected by $\mathcal{I}$ symmetry in the point group $\overline{1}$, and (a-2) in the rod geometry they appear along $z$ direction. (b) Type 2:~In the point group $\overline{3}$ the CHS on the sphere is invariant under $C_{3z}$ and $\mathcal{I}$ symmetries.
In point groups (c) Type 3:~$2/m$, $4/m$, $6/m$, (d) Type 4:~$mmm$, $m\overline{3}$, (e) Type 5:~$\overline{3}m$, (f) Type 6:~$4/mmm$, (g) Type 7:~$6/mmm$, (h) Type 8:~$m\overline{3}m$, the CHSs  form one, three, three, five, seven and nine great circles, respectively.} 
\label{mass}
\end{figure}

First, we consider positions of the CHSs in all the 92 space groups (SGs) with $\mathcal{I}$ symmetry or, equivalently the 92 type-I magnetic space groups (MSGs) with $\mathcal{I}$ symmetry.  
To this end, we consider a Dirac Hamiltonian on a surface of the system. 
In the simplest case, the surface Dirac Hamiltonian is written as 
\begin{equation}
\mathcal{H}(\boldsymbol{k})=\lambda (k_{1}\sigma_{y}-k_{2}\sigma_{x})+m\sigma_{z},
\end{equation}
where $\boldsymbol{\sigma}=(\sigma_{x},\sigma_{y},\sigma_{z})$
 are Pauli matrices, $\tilde{\boldsymbol{k}}=(k_{1},k_{2})$ is a wave-vector along the surface, and $\lambda$ and $m$ are real constants. On every surface, we define the mass $m$ for the surface Dirac Hamiltonian, whose general form is discussed in Appendix \ref{appendixa1}.
We first assume that the surface Dirac mass $m_{\boldsymbol{n}}$ is determined by the surface orientation, where $\boldsymbol{n}$ is the normal vector of the surface. We show the dependence of $m_{\boldsymbol{n}}$ on $\boldsymbol{n}$ in a system with a shape of a sphere, with a point on the sphere identified as a surface normal vector $\bs{n}$ (Fig.~\ref{mass}).    
In the $\mathcal{I}$ symmetric HOTIs without $\mathcal{T}$ symmetry, the sign of the mass term is reversed by $\mathcal{I}$ operation, $i.e.$ $m_{-\boldsymbol{n}}=-m_{\boldsymbol{n}}$ \cite{PhysRevB.97.205136}. Thus, the domain wall with $m_{\boldsymbol{n}}=0$ appears in a $\mathcal{I}$-symmetric manner, and on this $m_{\boldsymbol{n}}=0$ line, the CHS appears in a clockwise direction around the region with positive $m_{\boldsymbol{n}}$ (see Fig.~\ref{mass}(a-1)). Therefore, in the rod geometry along the $z$ direction, the CHSs follow from the distribution of $m_{\boldsymbol{n}}$ on the equator of the sphere (Fig.~\ref{mass}(a-2)).

Next, we consider the cases with $n$-fold rotational ($C_{nz}$) symmetry around the $z$-axis in addition to $\mathcal{I}$ symmetry. In the presence of $\mathcal{I}$ and $C_{3z}$ symmetries, $i.e.$ the point group $\overline{3}$, the CHS is invariant under $C_{3z}$ symmetry (Fig.~\ref{mass}(b)).
In the cases with $C_{2z}$ symmetry, the CHS on the sphere is
always a great circle on the mirror $xy$ plane due to mirror symmetry $M_{z}$ (Fig.~\ref{mass}(c)). 
In this case, chiral surface states protected by mirror symmetry appear on the side surfaces in the rod geometry 
along the $z$-axis. Meanwhile, if all the surfaces of the crystal are not mirror symmetric, CHSs appear. Furthermore, in the cases with $C_{4z}$ or $C_{6z}$ symmetries, the distributions of the mass terms are the  same as those with $C_{2z}$ symmetry. 
Thus, the CHS on the sphere is along a great circle in the point groups $2/m$, $4/m$ and $6/m$.
Similarly, in other point groups, the patterns of CHSs are classified into the five patterns (Fig.~\ref{mass}(d-h)), where the CHSs are located on great circles on the mirror planes. As shown in Appendix~\ref{ap:mirrorchern}, the sum of the mirror Chern numbers on $k_{z}=0$ and $k_{z}=\pi$ planes, is a non-zero odd number when $\mu_{1}=2$. Thus, the system with mirror symmetry has gapless surface modes protected by mirror symmetry when $\mu_{1}=2$, which makes the mass to vanish on the intersection between the sphere and the mirror plane.

 Notably, while on a mirror-invariant surface, the surface states are helical due to the nontrivial mirror Chern number, the hinge states predicted here are chiral as shown in Fig.~\ref{mass}(c-h). It is not a contradiction. For example, let us consider a 3D system of an $\mathcal{I}$-symmetric alternate stacking of Chern insulator layers with Chern numbers $\pm 1$ along the $z$-direction, with each layer being $M_{z}$ invariant. Even with $M_{z}$-invariant interlayer coupling, there appear gapless helical surface states on the $(100)$ surface due to a nontrivial mirror Chern number. On the other hand, if we cut out a finite crystal with $\mathcal{I}$ symmetry, the crystal consists of $N+1$ layers with the Chern number $+1$ and $N$ layers with the Chern number $-1$ or vice versa to preserve $\mathcal{I}$ symmetry. Thus, on the (100) surface, we should consider the $N\rightarrow \infty$ limit, and the surface states are helical. On the other hand, for an $\mathcal{I}$-symmetric crystal with a finite $N$ having no mirror-symmetric surfaces, by moving states of opposite chirality towards each other without breaking $\mathcal{I}$ symmetry, the chiral edge states hybridize to result in a non-vanishing Chern number $(N+1)\cdot 1 + N\cdot (-1)=1$, leading to CHSs. Thus, the massless line on the mirror planes in Fig.~\ref{mass}(c-h) corresponds to both helical surface states and CHSs.

This approach of the mass term is used in the previous works\cite{PhysRevX.8.031070, fang2019new} to classify the topological phases of topological crystalline insulators with $\mathcal{T}$ symmetry. Therefore, these previous works are not intended to list possible positions of hinge states. In contrast, in this paper, we focus on the position of CHSs in the system without $\mathcal{T}$ symmetry, not on the classification of topological phases. As we discuss in Appendix~\ref{appendixa1}, in the system without $\mathcal{T}$ symmetry, the transformation property of the mass term under rotation operations is different from the case with $\mathcal{T}$ symmetry, which gives crucial difference in the positions of CHSs.

\begin{figure}
\includegraphics[width=1\columnwidth]{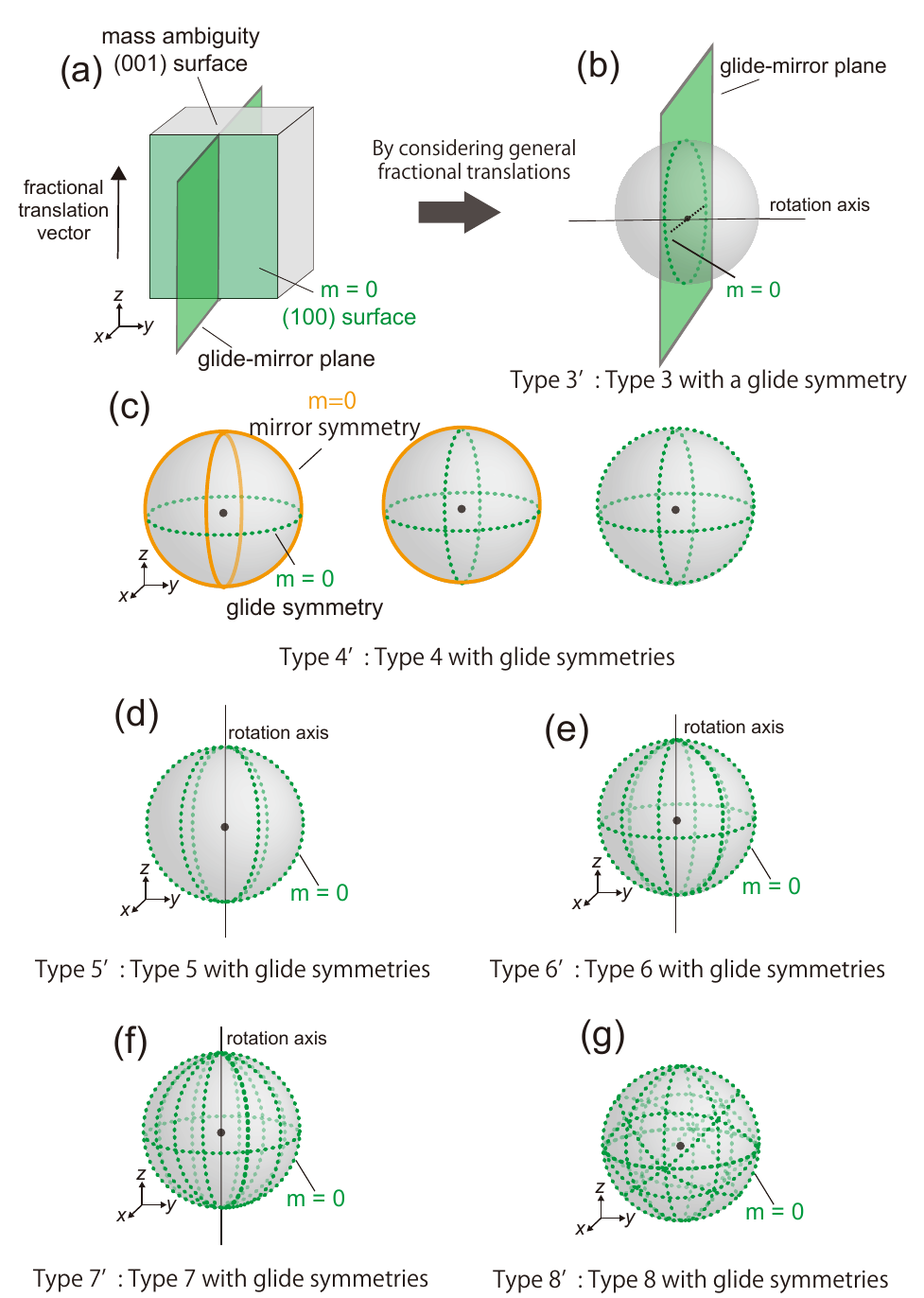}
\caption{The distributions of the mass term with glide symmetries. (a) A glide symmetry imposes glide-invariant surfaces such as $(100)$ to be gapless. On the other hand, the mass sign on the $(001)$ surface is indeterminate. (b) Within the glide mirror plane, the mass on the $(\alpha 0 \gamma)$ surface is zero when $\alpha \equiv 1$ (mod 2) and $\gamma \equiv 0$ (mod 2), while the mass is indeterminate when $\gamma \equiv 1$ (mod 2). Thus, on the sphere, along the great circle (dotted line) on the glide-mirror plane, the mass is not a continuous function of the surface normal vector. (b-f) We call the cases where the mirror planes in Types 3-8 are replaced with glide-mirror planes as Types $3'$-$8'$. (c) Type $4'$ contains several patterns which of the mirror symmetries in Type 4 are replaced by glide symmetries.} 
\label{105glidemass}
\end{figure}

Next, we consider the cases with glide symmetry. In systems with glide symmetry, the glide-$Z_{2}$ topological invariant is defined\cite{PhysRevB.91.161105, PhysRevB.91.155120}. Here, the question is a relationship between the glide-$Z_{2}$ invariant and the inversion-$\mathbb{Z}_{4}$ invariant. To see this, we consider minimal SGs which contain $\mathcal{I}$ and glide symmetries. These are three such SGs, No.~13, 14 and 15. The SGs No.~13 and 14 are for the simple (primitive) lattice, with the inversion center inside the glide-mirror plane in SG No.~13 and outside it in SG No.~14. The SG No.~15 is for the  base-centered lattice. It is found that in SG No.~13 and 15, a half of the inversion-$\mathbb{Z}_{4}$ invariant $\mu_{1}$ is equal to the glide-$Z_{2}$ invariant\cite{PhysRevB.100.165202, kim2020glide}, while in SG No.~14, it is equal to a sum of the glide-$Z_{2}$ invariant and a half of the Chern number, where the Chern number is even\cite{PhysRevB.100.165202}. In either case, because we are considering cases with a vanishing Chern number, a nontrivial inversion-$\mathbb{Z}_{4}$ invariant $\mu_{1}=2$ automatically means a nontrivial glide-$Z_{2}$ invariant.
Thus in our case of HOTIs, on the surfaces preserving the glide symmetry, there should appear topological gapless states, and the mass is zero.

To study the dependence of the surface mass on the surface orientation, we consider a system with a glide symmetry $\{M_{y}|00 \frac{1}{2}\}$. In this case, on the glide-invariant surfaces, which are perpendicular to the glide-mirror plane and along the direction of a fractional translation of the glide operations, the gap is zero in the HOTIs because of the nontrivial glide-$Z_{2}$ invariant (Fig.~\ref{105glidemass}(a)). More generally, by combining integer translations in the $x$, $y$ and $z$ directions with this glide operation $\{M_{y}|00 \frac{1}{2}\}$, it follows that $\{M_{y}|a, b, c+\frac{1}{2}\}$ is also among the symmetry operations of the system, where $a, b, c$ are integers. In this case, the Miller index of the corresponding glide-invariant surface is $(2c+1, 0, -2a)$, and the surface Dirac mass on this surface is zero. Therefore, on the $(\alpha, \beta, \gamma)$ surface with $\beta=0$, $\alpha \equiv 1$ (mod 2), and $\gamma \equiv 0$ (mod 2), the Dirac mass is zero.

Here, we note the following point:~so far we assume that the sign of $m_{\boldsymbol{n}}$ is uniquely determined by the surface orientation, but the assumption is violated when the system has glide symmetries. For example, we consider a glide symmetry $\{M_{y}|00 \frac{1}{2}\}$ once again. Suppose the surface mass term $m$ for the (001) surface $z=0$ is positive. Then from the glide symmetry, the (001) surface $z=1/2$ has a negative mass. Then the glide symmetry requires two terminations for the (001) surfaces to have opposite mass signs, violating our assumption.
Other than the (001) surface, there are surfaces with such ambiguity of the mass sign. In general, the $(\alpha, 0,  \gamma)$ surfaces have such ambiguity if $\gamma$ is an odd number, where $\alpha$ and $\gamma$ are integers, with its details in Appendix~\ref{ap_subsec_glide}. Thus, there are an infinite number of surface orientations on which the mass ambiguity arises. To summarize, when the surface normal is within the glide-mirror plane on the sphere shown in Fig.~\ref{105glidemass}(b), two cases for the surface mass coexist: the mass on the $(\alpha, 0, \gamma)$ surface is zero when $\alpha \equiv 1$ (mod 2) and $\gamma \equiv 0$ (mod 2), while the mass is indeterminate when $\gamma \equiv 1$ (mod 2). Thus, the mass with
the surface normal $\bs{n}$ within the glide mirror-plane sensitively depends on the surface orientation, and it is no longer a smooth function of $\bs{n}$. In the mass distribution on the sphere, such $\bs{n}$ forms a great circle on the glide-mirror plane. We show this great circle as a broken circle, in contrast with the solid circle on the mirror plane, where the mass is zero. We call the cases where the mirror planes in Types 3-8 are replaced with a glide mirror plane as Types $3'$-$8'$ (Figs.~\ref{105glidemass}(b-g)).

Next, we consider the cases with screw symmetries. The combination between $\mathcal{I}$ symmetry and screw $\{C_{2z}|00\frac{1}{2}\}$ symmetry leads to the mirror symmetry $\{M_{z}|00\frac{1}{2} \}$, and the mirror Chern number is nontrivial when the Chern number on $k_{z}=0$ plane is zero and $\mu_{1}=2$ as discussed in Appendix \ref{sec:appnedixE:connectionbetweenmirror}. It leads to the emergence of the surface states protected by mirror symmetry (Type 3 in Fig.~\ref{mass}).

\begin{table}
\setlength{\extrarowheight}{.7ex}
\begin{center}
\caption{The patterns of the CHSs on the sphere for 92 type-I MSGs with $\mathcal{I}$ symmetry. 
The distributions of the CHSs are classified into 14 types: the 8 types (Types 1-8) shown in Fig.~\ref{mass} (a-1, b$-$h) and the 6 types with glide symmetries (Types $3'$-$8'$) shown in Fig.~\ref{105glidemass}(b-g). PG denotes the name of the point group. In Types 1-8, the MSGs do not include glide symmetries. On the other hand, in Types $3'$-$8'$, the MSGs include glide symmetries, and these cases correspond to the mass distributions which are obtained by replacing the mirror planes in Types 3-8 with glide-mirror planes.  \label{tab:allpattern_hinge}}
\begin{tabular}{|c|c|c|}
\hline
\hline
PG &~Type-I MSG~&~Type of CHS~\\\hline 
$\overline{1}$  & 2. & Type 1  \\
\hline
$\overline{3}$  & 147, 148. & Type 2\\
\hline 
\multirow{2}{*}{$2/m$}  &  10, 11, 12. &  Type 3 \\
\cline{2-3}
   &13, 14, 15.&  Type $3'$ \\
\hline
\multirow{2}{*}{$4/m$}  & 83, 84, 87.& Type 3\\
\cline{2-3}
  & 85, 86, 88.& Type $3'$\\
  \hline
$6/m$ &  175, 176. & Type 3 \\
\hline
\multirow{4}{*}{$mmm$}  & 47, 65, 69, 71.  & Type 4\\
\cline{2-3}
 & 48, 49, 50, 51, 52, 53, 54, 55, & \multirow{3}{*}{Type $4'$}\\
 & 56, 57, 58, 59, 60, 61, 62, 63,  & \\
 & 64, 66, 67, 68, 70, 72, 73, 74.  & \\
 \hline
\multirow{2}{*}{$m\overline{3}$} & 200, 202, 204. & Type 4\\
\cline{2-3}
 & 201, 203, 205, 206. & Type $4'$\\
\hline
\multirow{2}{*}{$\overline{3}m$}  & 162, 164, 166. & Type 5\\
\cline{2-3}
  & 163, 165, 167. & Type $5'$\\
\hline 
\multirow{4}{*}{$4/mmm$}   & 123, 139.  & Type 6\\
\cline{2-3}
   & 124, 125, 126, 127, 128, 129, 130,   & \multirow{3}{*}{Type $6'$}\\
   & 131, 132, 133, 134, 135, 136, 137,   & \\
   & 138, 140, 141, 142. & \\
\hline
\multirow{2}{*}{$6/mmm$}  & 191. & Type 7\\
\cline{2-3}
& 192, 193, 194. & Type $7'$\\
\hline
\multirow{2}{*}{$m\overline{3}m$} & 221, 225, 229. & Type 8\\
\cline{2-3}
& 222, 223, 224, 226, 227, 228, 230. & Type $8'$\\
\hline
\hline
\end{tabular}
\end{center}
\end{table}

In this way, we identified all the 14 possible patterns of the mass distribution on the sphere and the resulting positions of CHSs (Types 1-8 in Fig.~\ref{mass}(a-h) and Types $3'$-$8'$ with glide symmetries in Fig.~\ref{105glidemass}(b-g)) for all the 92 type-I MSGs with $\mathcal{I}$ symmetry as shown in Table~\ref{tab:allpattern_hinge}. We obtain this result using an observation that $m_{\boldsymbol{n}}$ behaves as a pseudoscalar with its details in Appendix \ref{asymm_appendixa}. Here, we note the meaning of the spherical figures in Figs.~\ref{mass} and \ref{105glidemass}. They represent dependence of the surface Dirac mass on the surface normal $\bs{n}$. Therefore, for a given crystal shape with various facets, we apply one of the 14 patterns in Figs.~\ref{mass} and \ref{105glidemass} to find the signs of the individual crystal facets, and then the positions of the CHSs are determined as domain walls for the sign of the surface mass. Hence, depending on crystal shapes, the positions of CHSs may look different from Figs.~\ref{mass} and \ref{105glidemass}.

\section{Chiral hinge states in antiferromagnetic higher-order topological insulators}\label{sec:hingestatesinlayered}

Based on the mass distribution $m_{\boldsymbol{n}}$ dependent on the surface normal vector $\boldsymbol{n}$, one can determine positions of CHSs for a given crystal shape.
In Sec.~\ref{sec2:chiral_antiferro}, we find an interesting possibility in HOTIs with glide symmetry. Namely, for particular choices of surface orientations, the surface Dirac mass changes its sign, when the surface termination is shifted by a half of the lattice constant. Since the mass distribution as a function of $\bs{n}$ determines the positions of CHSs, our finding in Sec.~\ref{sec2:chiral_antiferro} suggests that in such cases, a change in the surface termination, while keeping the surface orientation specified by $\bs{n}$, will change the positions of the CHSs.
To show this, we consider an example of a glide-symmetric system in cylindrical geometries along the direction of the fractional translation of glide symmetry, and we choose the Types $3'$ and $5'$ in Fig.~\ref{105glidemass} as an example.
 
\begin{figure}
\includegraphics[width=0.95\columnwidth]{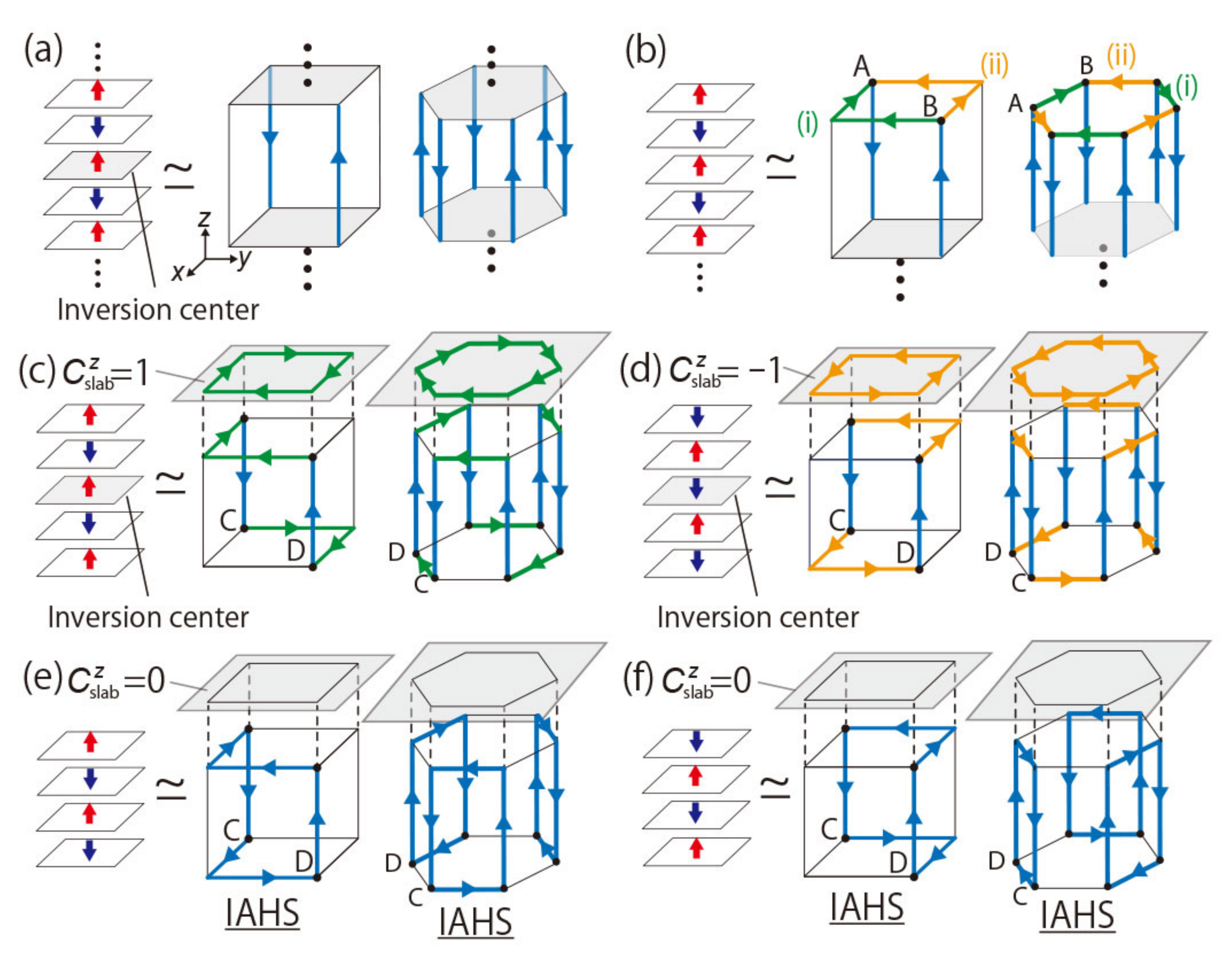}
\caption{Positions of the CHSs and the slab Chern number $C^{z}_{\rm slab}$. (a) The CHSs appear along the $z$ direction, $i.e.$ the stacking direction in the rod geometries parallel to the fractional translation vector in the glide symmetry in  Type $3'$ and Type $5'$. (b) The number of incoming hinge modes should be equal to that of outgoing hinge modes at the corners A and B. There are two possibilities for CHSs, (i) and (ii) in both cases corresponding to Types $3'$ and $5'$. (c-f) There are four possible patterns of the positions of CHSs. The cases with $N={\rm odd}$ are in (c) $C^{z}_{\rm slab}=1$ and (d) $C^{z}_{\rm slab}=-1$. The cases with $N={\rm even}$ are in (e) and (f), and IAHS appear in these cases. }
\label{hinge}
\end{figure}

Here, we consider crystals in Type $3'$ in a shape of a parallelepiped and in Type $5'$ in a shape of a hexagonal prism (Fig.~\ref{hinge}(a)).
$\mathcal{I}$-symmetric 3D layered antiferromagnetic HOTIs with $\mu_{1}=2$.
Here, the staggered magnetization is along the $z$-axis, which is along the stacking direction.
In this section, we show that the positions of the CHSs depend on the parity of the number of layers $N$, and IAHS generally appear in the cases with even $N$, where $\mathcal{I}$ symmetry is broken.
We assume that each layer has $\mathcal{I}$ symmetry.  
First, we consider CHSs along the direction of the stacking, $i.e.$~the $z$ direction as shown in Fig.~\ref{hinge}(a) which follow from Fig.~\ref{mass}.
Here we assume that the number of layers is much larger than the penetration depth of the CHSs.
In this case, CHSs along the stacking direction always emerge irrespective of the number of layers because of the topological property in the bulk, $i.e.$ $\mu_{1}=2$ \cite{PhysRevB.101.115120}.

Next, we consider behaviors of CHSs at the corners A and B in Fig.~\ref{hinge}(b). At each corner, the number of incoming hinge modes should be equal to that of outgoing hinge modes because otherwise a charge will be accumulated at a corner in proportion with time, due to the fact that each hinge mode provides one-channel transport.
This argument is similar to the one in Ref.~[\onlinecite{berrypahaseinelectronic}] applied for chiral edge currents in a two-dimensional (2D) insulating ferromagnet.
Therefore, CHSs should appear either along (i) or along (ii) between the corners A and B (Fig.~\ref{hinge}(b)). 
In addition, a similar discussion is applicable to the corners C and D. Therefore there are four possible patterns of the positions of CHSs as shown in Figs.~\ref{hinge}(c-f). In Figs.~\ref{hinge}(c) and~\ref{hinge}(d), the positions of CHSs are $\mathcal{I}$ symmetric. 

\subsection{Cases with an odd number of layers}
First of all, we consider the cases with odd $N$ with the magnetization of each layer given by $\uparrow \downarrow \cdots \downarrow \uparrow$, which  represents the magnetizations of the individual layers from top to bottom. 
When $N$ is odd, the system has $\mathcal{I}$ symmetry, and such an $\mathcal{I}$-symmetric 2D slab of a 3D HOTI with $\mathcal{I}$ symmetry is shown to be a 2D Chern insulator with the Chern number $C^{z}_{\rm slab}\equiv 1$ (mod 2)\cite{PhysRevB.98.205129, PhysRevResearch.2.013300, wieder2018axioninsulatorpump}. 
Here the Chern number $C^{z}_{\rm slab}$ in 2D systems is defined by
\begin{equation}
C^{z}_{\rm slab}=\frac{1}{2\pi} \int_{\rm BZ}  dk_{x}dk_{y} {\rm Tr}[\mathcal{F}_{xy}(\boldsymbol{k})].
\end{equation}
Here 
$
\mathcal{F}_{ab}(\boldsymbol{k})=\partial_{a} \mathcal{A}_{b} (\boldsymbol{k}) - \partial_{b} \mathcal{A}_{a}(\boldsymbol{k})
$
is the Berry curvature written in terms of the Berry connection of the occupied bands, $(\mathcal{A}_{\mu}(\boldsymbol{k}))^{\alpha \beta}=i \bra{u_{\alpha}}\partial_{\mu}\ket{u_{\beta}}$.
In the present case of the layered antiferromagnet, when $N={\rm odd}$, $\mathcal{I}$ symmetry is preserved, and the slab Chern number $C^{z}_{\rm slab}$ (mod 2) satisfies \cite{PhysRevResearch.2.013300}; 
\begin{equation}\label{chernmu1}
C^{z}_{\rm slab}\equiv \frac{1}{2}\mu_{1} \equiv 1 \ \ ({\rm mod}\ 2),
\end{equation}
since we are considering HOTIs with $\mu_{1}\equiv 2$ (mod 4).
In particular, for odd $N$ with $\uparrow \downarrow \cdots \downarrow \uparrow$ and $\downarrow \uparrow \cdots \downarrow$, we can choose $C^{z}_{\rm slab}=1$ and $C^{z}_{\rm slab}=-1$ respectively without losing generality (see Appendix~\ref{appendixc:chernnumber}), leading to the positions of CHSs shown in Figs.~\ref{hinge}(c) and \ref{hinge}(d).

\subsection{Cases with an even number of layers}Next, when $N$ is even, $\mathcal{I}$ symmetry is broken and $C^{z}_{\rm slab}\equiv 1$ (mod 2) does not hold.  For even $N$ with $\uparrow \downarrow \uparrow \downarrow \cdots \downarrow$, it can be understood as $\uparrow \downarrow \cdots \uparrow + \downarrow \uparrow \cdots \downarrow$, $i.e.$ the composition of odd-$N$ layers with $C^{z}_{\rm slab}=1$ and odd-$N$ layers with $C^{z}_{\rm slab}=-1$. The resulting positions of CHSs are shown in Fig.~\ref{hinge}(e) because two counter-propagating CHSs on the same hinge will hybridize and open a gap. Likewise, for even $N$ with $\downarrow \uparrow \cdots \uparrow$, the CHSs are as shown in Fig.~\ref{hinge}(f), which is obtained from Fig.~\ref{hinge}(e) via an $\mathcal{I}$ operation. These positions of CHSs are not $\mathcal{I}$ symmetric; namely,  they are IAHS. Thus, while the  CHS forms a single loop when $N={\rm odd}$ (see Figs.~\ref{hinge}(c) and \ref{hinge}(d)), those with $C_{3z}$ symmetry when $N$ = even do not form a single loop, but three loops instead (Figs.~\ref{hinge}(e) and \ref{hinge}(f)). The difference in the positions of CHSs can be observed by transport measurements on the hinges of crystals.

Interestingly, though a slab with even $N$ does not preserve $\mathcal{I}$ symmetry, it has IAHS due to bulk topology protected by bulk $\mathcal{I}$ symmetry. It is also interesting that the positions of the IAHS for even $N$ are also uniquely determined in this case, and they are different from those for odd $N$.
EuIn$_{2}$As$_{2}$ with even $N$  possesses $\mathcal{I}\tilde{\mathcal{T}}$ symmetry where $\tilde{\mathcal{T}}\equiv C_{2z}\mathcal{T}$, from which the Chern number is 
\begin{equation}
C^{z}_{\rm slab}=0,
\end{equation}
 when $N$ is even as proved in Appendix \ref{appnedixb:chernzero}.

\section{Model calculations}\label{sec:modelcalculations}

\begin{figure}[t]
\includegraphics[width=1\columnwidth]{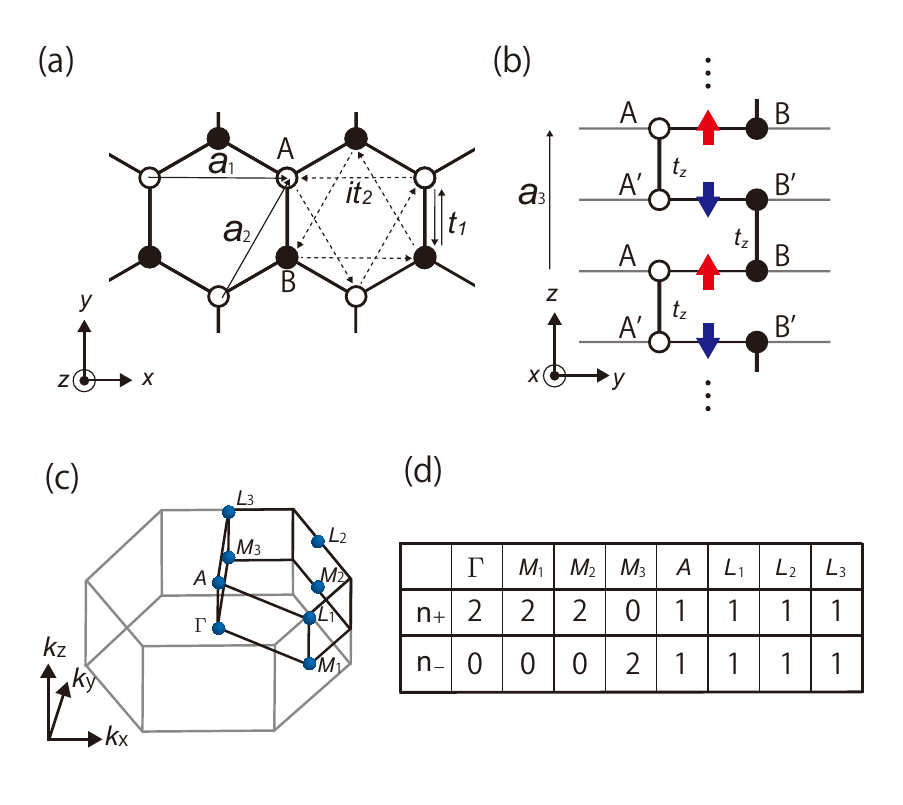}
\caption{\label{model_calc}Tight-binding model of the layered AFM. (a) Each layer forming the Haldane model on  the honeycomb lattice with the A and B sublattices. Primitive translation vectors are $\boldsymbol{a}_{1}=(a,0,0)^{T}$ and $\boldsymbol{a}_{2}=(a/2,\sqrt{3}a/2,0)^{T}$, where $a$ is a lattice constant. 
(b) Schematic picture of the interlayer hopping. Each layer can be regarded as a ferromagnetic system, and the magnetizations of each layer change alternately.  From this, the whole system has four sublattices, A, B, A' and B' sublattices.   
The interlayer hopping between A sites alternates with that between B sites. $\boldsymbol{a}_{3}=(0,0,2a)^{T}$ is the translation vector along the $z$ direction.
(c) The Brillouin zone of this model, and there are eight TRIM. (d) The number of occupied states with even and odd parities at the eight TRIM. 
}
\end{figure}

In this section, we use a tight-binding model of a layered AFM showing a HOTI phase to study the behaviors of the IAHS.
This model was proposed as a model of a HOTI with CHSs in Ref.~[\onlinecite{PhysRevB.98.245102}]. 
This model is an alternate stacking of layers of the Haldane model within the $xy$ plane with the Chern number $=\pm 1$ \cite{PhysRevLett.61.2015}, and their magnetizations alternate correspondingly (Figs.~\ref{model_calc}(a) and \ref{model_calc}(b)).
The Haldane model is a tight-binding model on the honeycomb lattice representing a ferromagnet, and its Hamiltonian within the $\alpha$-th layers is written as 
\begin{equation}
H_{xy}=t_{1}\sum_{\langle ij \rangle, \alpha}c^{\dagger}_{i\alpha}c_{j\alpha}+it_{2}\sum_{\langle \langle ij \rangle \rangle, \alpha}(-1)^{\alpha} \nu_{ij}c^{\dagger}_{i\alpha}c_{j\alpha},
\end{equation}
 where $i$ and $j$ run over the sites in the layer $\alpha$, $t_{1}$ is the hopping strength for the first-neighbor pairs $\langle ij \rangle$, $t_{2}$ is that for the second-neighbor pairs, and $\nu_{ij}=1$ ($\nu_{ij}=-1$) if the second-neighbor hopping path is counterclockwise (clockwise) in the hexagonal plaquette. 
 The Fermi energy is set to be $E_{F}=0$.
 We assume $t_{1}$ and $t_{2}$ to be negative; then each layer is a Chern insulator with the Chern number $C=(-1)^{\alpha+1}$. 
 Therefore, the Chern numbers of the individual layers change alternately, and so does their magnetization. Therefore, the model is a layered AFM. Next we introduce a hopping $t_{z}$ between the layers as shown in Fig.~\ref{model_calc}(b), and this Hamiltonian is given by 
\begin{align}
H_{z}=&
\frac{t_{z}}{2}\sum_{i \in A, \alpha}(1-(-1)^{\alpha})c^{\dagger}_{i \alpha}c_{i \alpha+1} \nonumber \\
&\ \ \ \ \ \ +\frac{t_{z}}{2}\sum_{i \in B, \alpha}(1+(-1)^{\alpha})c^{\dagger}_{i \alpha}c_{i \alpha+1}+{\rm h.c.}
\end{align}
The overall Hamiltonian is $H=H_{xy}+H_{z}$, and it has $\mathcal{I}$ symmetry in the bulk.
This interlayer coupling breaks $\mathcal{S}(=\mathcal{T}\tau_{\frac{1}{2}})$ symmetry but preserves bulk inversion symmetry, which leads to gapped surface states and gapless CHSs. Here, we assume that the interlayer coupling is weak
so that the gap remains open even when $t_{z}$ is continuously changed to zero. Then the topological properties of the layered system are the same as those of a stack of decoupled 2D Chern insulators.
Therefore, $C^{z}_{\rm slab}$ is obtained by the sum of the Chern number of each layer.

\begin{figure}[t]
\includegraphics[width=0.95\columnwidth]{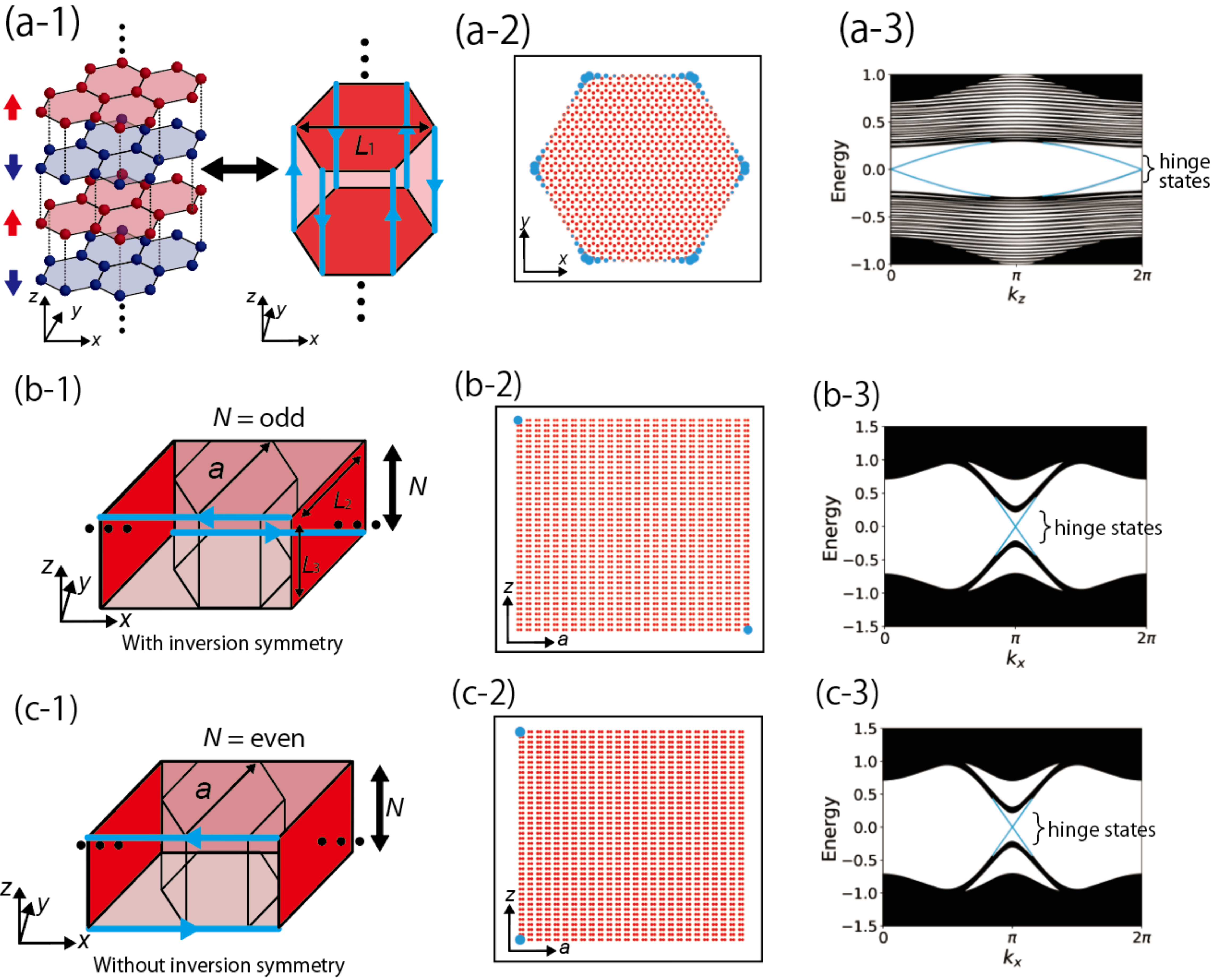}
\caption{\label{sup_model_calc2}The positions of CHSs and the band structures of the tight-binding model with parameters $t_{1}=-1$, $t_{2}=-0.2$ and $t_{z}=-0.3$. The real-space distributions of the CHSs in (a-2) (b-2) and (c-2) are shown as  the size of the blue dots. (a) Results for rod geometry along the $z$ direction. (a-1) Schematic diagram of the layered AFM comprising layers of the Haldane model, and the positions of CHSs for hexagonal rod geometry. (a-2) real-space distribution of zero-energy modes in the $xy$ plane. 
The longest diagonal length of the hexagonal system in the real space is $L_{1}=25a$. 
 (a-3) The band structure.
(b)(c) Results for rod geometry which is finite along the $z$ and $a_{2}$ directions, and infinite along the $a_{1}$ direction. The lengths of the system along the $a_{2}$ are $L_{2}=25a$ in both (b) and (c), and those along the $a_{3}$ are $L_{3}=40a$ ($N=41$) in (b) and $L_{3}=41a$ ($N=42$) in (c).
The numbers of the layers are odd in (b) and even in (c). (b-1) and (c-1) are schematic diagrams of CHSs.  (b-2) and (c-2) are the real-space distribution of zero-energy modes in the red plane in (b-1) and (c-1). (b-3) and (c-3) The band structures.
When $N={\rm odd}$, the positions of CHSs are $\mathcal{I}$ symmetric corresponding to Fig.~\ref{hinge}(c). 
When $N={\rm even}$, the positions of CHSs $i.e.$ IAHS are not $\mathcal{I}$ symmetric corresponding to Fig.~\ref{hinge}(e). These model calculations were performed using the PythTB package \cite{pythTB}.
}
\end{figure}

In 3D systems, there are eight time-reversal invariant momenta (TRIM) denoted by $\Gamma_j$. The topological phases are characterized by the $\mathbb{Z}_{4}$ index for $\mathcal{I}$ symmetric systems in class A, one of Altland-Zirnbauer symmetry classes \cite{PhysRevB.55.1142}. 
The $\mathbb{Z}_{4}$ index is defined as  \cite{po2017symmetry}
\begin{equation}
\mu_{1}\equiv \frac{1}{2}\sum_{\Gamma_{j}: {\rm TRIM}} [n_{+}(\Gamma_{j})-n_{-}(\Gamma_j)]\ \ \ ({\rm mod}\ 4), 
\end{equation}
where $n_{\pm}(\Gamma_{j})$ are the number of the occupied states with even- and odd-parity eigenvalues at the TRIM $\Gamma_{j}$, respectively.
The eight TRIM are shown in Fig.~\ref{model_calc}(c), and the parity eigenvalues at each TRIM are shown in Fig.~\ref{model_calc}(d). From these parity eigenvalues, the $\mathbb{Z}_{4}$ topological index is $\mu_{1}=2$ in this model, which leads to HOTIs with CHSs \cite{PhysRevB.101.115120}.

\begin{figure}[t]
\includegraphics[width=0.95\columnwidth]{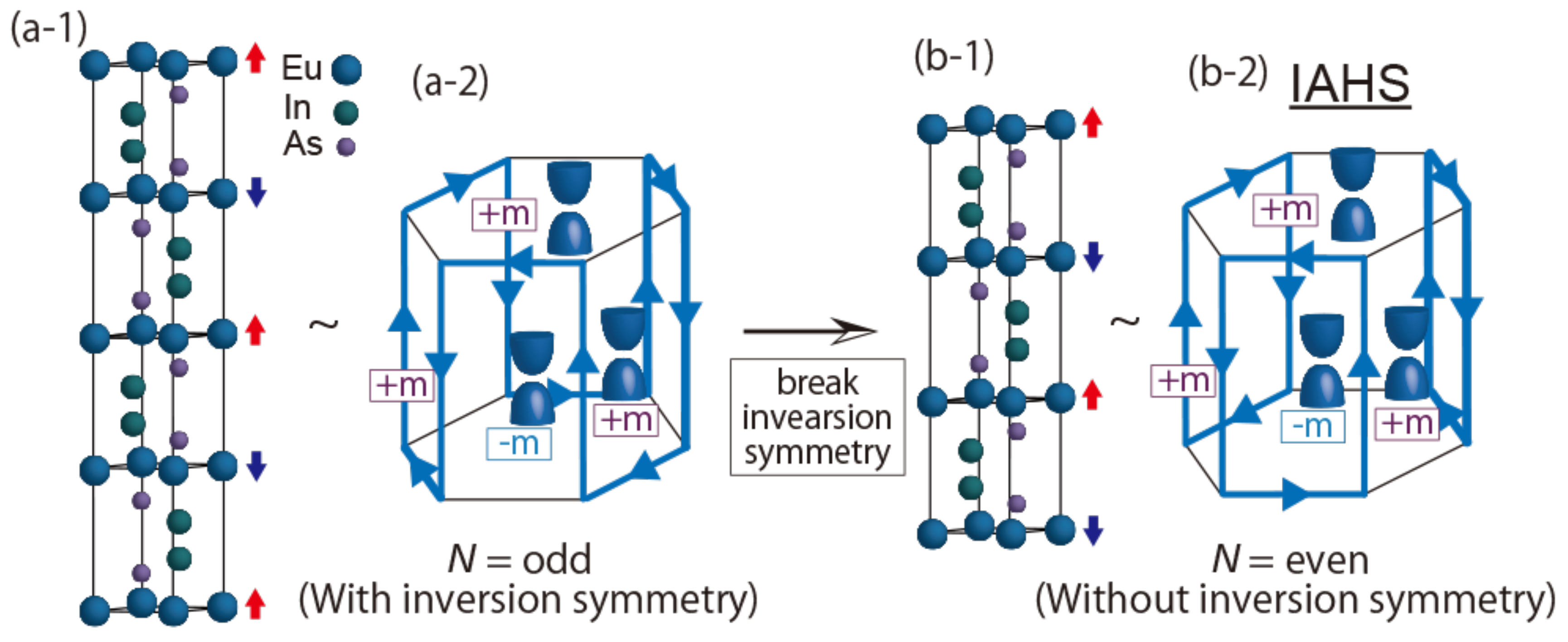}
\caption{\label{fig:material}Inversion-symmetric and inversion-asymmetric hinge states in EuIn$_{2}$As$_{2}$. (a-1) The crystal with odd $N$ for EuIn$_{2}$As$_{2}$. (a-2) CHSs corresponding to the case with odd $N$.  (b-1) The crystal with even $N$ for EuIn$_{2}$As$_{2}$. (b-2) IAHS with $C_{3z}$ symmetry in EuIn$_{2}$As$_{2}$ with even $N$.} 
\end{figure}

In the following, we perform calculations for the tight-binding model using the PythTB package \cite{pythTB}.
We calculate the band structure with the rod geometry along the $z$ direction, and the result is  shown in Fig.~\ref{sup_model_calc2}(a). This model has $C_{3z}$ symmetry, and therefore the CHSs appear at the positions which are related by $C_{3z}$ symmetry as shown in Figs.~\ref{sup_model_calc2}(a-1) and \ref{sup_model_calc2}(a-2).  Next we calculate the band structure with the rod geometry which is finite  along the directions of $z$ axis and of the primitive translation vector $\boldsymbol{a_{2}}$, and infinite along the $\boldsymbol{a_{1}}$ direction. The results are shown in Figs.~\ref{sup_model_calc2}(b) and \ref{sup_model_calc2}(c), for $N={\rm odd}$ and $N={\rm even}$ respectively. 
When $N={\rm odd}$ (Fig.~\ref{sup_model_calc2}(b-2)), the system is $\mathcal{I}$ symmetric, and then the positions of CHSs are $\mathcal{I}$ symmetric, in agreement with Fig.~\ref{hinge}(c) with $C^{z}_{\rm slab}=1$. When $N={\rm even}$, the system is not $\mathcal{I}$ symmetric, and the positions of CHSs are as shown in Fig.~\ref{sup_model_calc2}(c-2), that is, IAHSs appear with $C^{z}_{\rm slab}=0$.

The MSG of the tight-binding model is $P6'_{3}/m'm'c$ $(\# 194. 268)$. This MSG contains two kinds of symmetry operations:~a glide symmetry $\{ M_{y}|00\frac{1}{2} \}$, and a combination of screw and $\mathcal{T}$ operations, $\mathcal{T} \{C_{2z}|00\frac{1}{2}\}$, both of which lead to a sign inversion of the surface Dirac mass on the (001) surface. The MSG $P6'_{3}/m' m' c$ ($\# 194.268$) has a maximum subgroup type-I MSG No.~167. 
According to Table \ref{tab:allpattern_hinge}, it is similar to the point group $\overline{3}m$ with three glide symmetries (Type $5'$ in Fig.~\ref{105glidemass}).

 According to Ref.~[\onlinecite{PhysRevLett.122.256402}], EuIn$_{2}$As$_{2}$ is a layered antiferromagnetic AXI with the magnetic moment along $z$ axis (Fig.~\ref{fig:material}(a-1)). The MSG of EuIn$_{2}$As$_{2}$ is $P6'_{3}/m' m' c$ ($\# 194.268$), the same with our model. Thus, EuIn$_{2}$As$_{2}$ is an ideal material to study the emergence of IAHS. Remarkably, the mass term in EuIn$_{2}$As$_{2}$ will change its sign under $\mathcal{I}$, and invariant under $C_{3z}$, which makes the mass term having an alternate sign between adjacent side surfaces of the hexagonal crystal shown in Fig.~\ref{fig:material}(a-2) and Fig.~\ref{hinge}(c).  
Moreover, if $N$ is even, which breaks $\mathcal{I}$ symmetry (Fig.~\ref{fig:material}(b-1)), CHSs will exist in an $\mathcal{I}$-asymmetric configuration (Figs.~\ref{fig:material}(b-2) and \ref{hinge}(e)). 

In the AXIs, such as EuIn$_{2}$As$_{2}$, a quantized magnetoelectric effect is expected. For its observation, we need to attach electrodes so that they are electrically isolated from each other. The CHSs can short-circuit them, then obstructing the measurement. Our analysis shows that the case with $N={\mathrm{even}}$ may facilitate the measurement since the IAHS consist of multiple loops.

\section{Conclusion}\label{sec:conclusion}
In summary, we found all the possible configurations of CHSs in HOTIs in type-I MSGs with $\mathcal{I}$ symmetry through an analysis of the sign of the surface Dirac mass. The configurations are uniquely determined from each MSG as a massless line of the surface Dirac mass. Through this study, we found that in systems with glide symmetry, two surface termination with the same particular surface orientations have the opposite signs of the mass. This possibility leads to a drastic difference in positions of CHS in AXIs realized in layered AFMs between the cases with an even number of layers and an odd number of layers. In particular, in the case with an even number of layers, IAHS appear at the $\mathcal{I}$-asymmetric positions  because $\mathcal{I}$ symmetry is broken in the whole system. Nonetheless, IAHS are protected by $\mathcal{I}$ symmetry in the bulk, and they are characterized by the topological $\mathbb{Z}_{4}$ invariant. This difference in the positions of CHSs can be observed by transport measurements through hinges.  Furthermore, we predict that IAHS appear in an AXI~EuIn$_{2}$As$_{2}$  with an even number of layers.

\begin{acknowledgments}
This work was supported by Japan Society for the Promotion of Science (JSPS) KAKENHI Grants No.~JP18J23289, No.~JP18H03678, and No.~JP20H04633, and by the Ministry of Education, Culture, Sports, Science, and Technology Elements Strategy Initiative to Form Core Research Center (TIES), Grant Number JPMXP0112101001.
\end{acknowledgments}

\appendix\
\section{Surface theory of chiral hinge states in 92 type-I magnetic space groups with $\mathcal{I}$ symmetry}\label{asymm_appendixa}

\begin{figure*}
\centerline{\includegraphics[width=1.8\columnwidth]{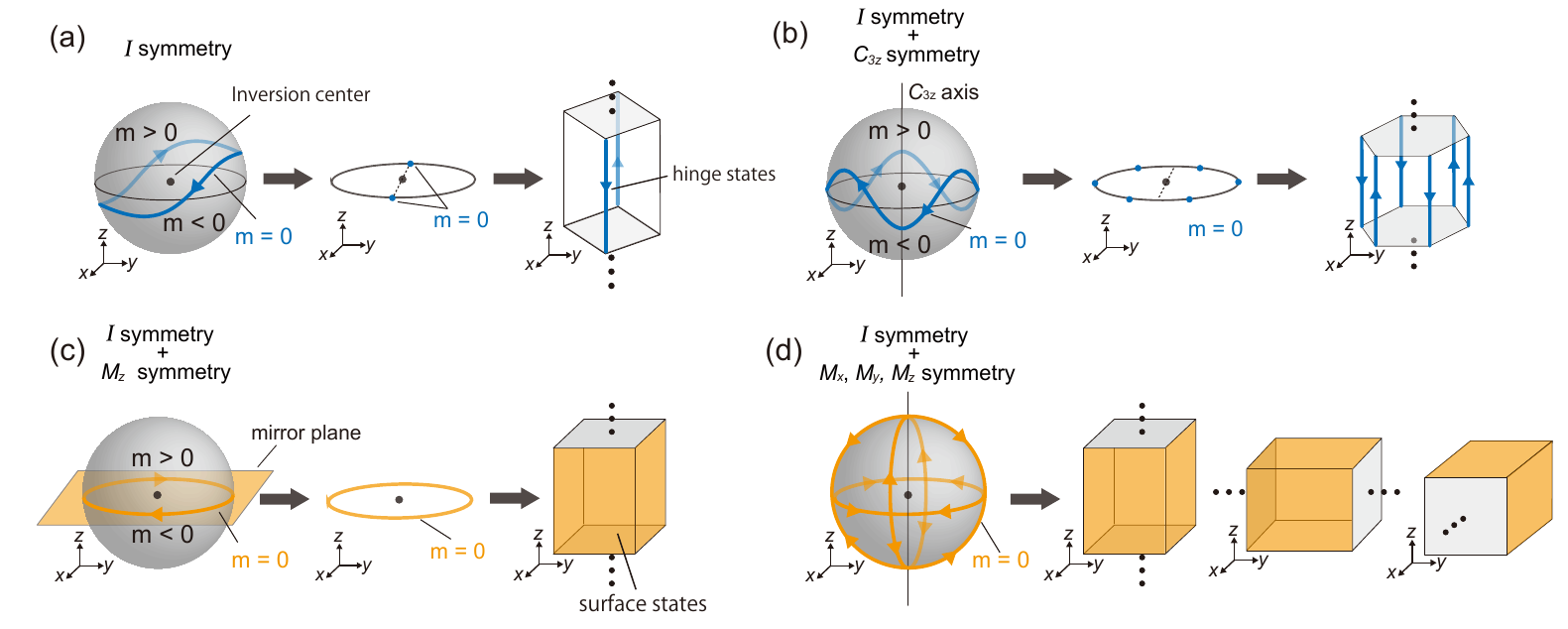}}
  \caption{The distributions of the mass term on a sphere and the positions of CHSs in the rod geometry. (a) The signs of the mass on the sphere with $\mathcal{I}$ symmetry. The mass sign on the equator of the sphere corresponds that on the surface in the rod geometry along the $z$ directions. (b) The case with $\mathcal{I}$ symmetry and $C_{3z}$ symmetries. (c) The case with $\mathcal{I}$ symmetry and mirror $M_{z}$ symmetry with respect to $xy$ plane, leading also to $C_{2z}$ symmetry. The chiral edge states flow along the great circle at the intersection between the sphere and the mirror plane. It leads to the surface states in the rod geometry along the $z$ direction. (d) In the case with $M_{x}$, $M_{y}$ and $M_{z}$ symmetries, chiral edge states appear along three great circles, and the surface states appear in the rod geometries along $x$, $y$ and $z$ directions.}
\label{ap_hinge_sphere}
\end{figure*}

In this appendix, we identified all the possible patterns of the CHSs for all the 92 type-I MSGs with $\mathcal{I}$ symmetry.

\subsection{Setup of the problem}\label{appendixa1}

First, we start from the surface theory of HOTIs, and from this information, we discuss emergence of CHSs. We assume that surfaces of HOTIs with $\mathcal{I}$ symmetry are represented by the Dirac Hamiltonian. In the simplest case, the surface Dirac Hamiltonian $\mathcal{H}(\boldsymbol{k})$ with a mass term is given by 
\begin{equation}
\mathcal{H}(k_{1},k_{2})=\lambda(k_{1}\sigma_{y}-k_{2}\sigma_{x})+m\sigma_{z},
\end{equation}
where $\sigma_{i}$ ($i=x,y,z$) represent the Pauli matrices, $(k_{1},k_{2})$ is the wave-vector along the surface, and $\lambda$ and $m$ are real constants.
Here, $m$ represents the surface Dirac mass. In fact, this form of the Hamiltonian is just an example, and in general cases, the Hamiltonian $\mathcal{H}(\boldsymbol{k})$ on the surface with a unit normal vector $\boldsymbol{n}$ can be represented by    
\begin{equation}
\mathcal{H}(k_{1},k_{2})=\sum_{i,j} A_{ij}\sigma_{i}k_{j},
\end{equation}
where $k_{3}\equiv m$ is a Dirac mass. In addition, $A_{ij}$ are real, and we can always assume ${\rm det}(A)>0$, because otherwise we change the definition of $m$ by $m \leftrightarrow -m$ so that ${\rm det}(A)$ becomes positive.
This process uniquely determines the sign of the mass. 
Furthermore, for the moment we assume that the sign of the mass term is determined uniquely by the surface normal vector $\boldsymbol{n}$.

Now we discuss a crucial difference between HOTIs with and without $\mathcal{T}$ symmetry. In HOTIs without $\mathcal{T}$ symmetry, the sign of the mass can be defined without ambiguity, because it is associated with the chiral direction of the CHS, $i.e.$ the CHS goes around the region with a positive mass in the clockwise way. In contrast, for helical hinge states in HOTIs with $\mathcal{T}$ symmetry, one cannot define their flow direction (because they contain flows in both directions), and one cannot define the sign of the mass $m_{\boldsymbol{n}}$ in a gauge-independent way.

In the following, we use the fact that the mass term in HOTIs with $\mathcal{I}$ symmetry is a pseudoscalar under point-group operations, namely, it changes sign under improper rotations such as $\mathcal{I}$ and mirror operations. 
This fact is justified from the following three properties of CHSs.

(i) It is established that the mass term in HOTIs with $\mu_{1}=2$ changes sign under $\mathcal{I}$ operation, and therefore the mass term satisfies $m_{-\boldsymbol{n}}=-m_{\boldsymbol{n}}$.\cite{PhysRevB.97.205136}
Therefore, the domain wall with $m_{\boldsymbol{n}}=0$ appears between the regions with a positive $m_{\boldsymbol{n}}$ and that with a negative $m_{\boldsymbol{n}}$ (see Figs.~\ref{ap_hinge_sphere}(a) and \ref{ap_hinge_sphere}(b)). Along this massless line, CHS exists in a clockwise way around the $m_{\boldsymbol{n}}>0$ region. 
In this way, CHSs appear along the domain wall with $m_{\boldsymbol{n}}=0$. On the other hand, in normal insulators ($\mu_{1}=0$), the sign of the mass term does not change under $\mathcal{I}$ operation.

(ii) Under the rotations, the sign of the mass term does not flip. 
One can see that it leads to a contradiction if the sign of the mass changes under rotation operations. For example, in a system with  $C_{2z}$ symmetry, if the sign of the mass term changes under $C_{2z}$ operation, the CHS appears along the domain wall passing through the north pole and south pole, which are $C_{2z}$ invariant points on the sphere. It obviously contradicts $C_{2z}$ symmetry because of chiral nature.  
Thus, the mass does not change under $C_{2z}$, and therefore the CHS does not follow only from the $C_{2z}$ symmetry. It is consistent with the trivial strong index of symmetry-based indicator\cite{PhysRevB.98.115150} in type-I MSG No.3. In this argument, the chiral nature of the hinge states is crucial.

(iii) The sign of the mass term flips under a mirror operation. From this, the gapless modes appear on the mirror plane, which are identified as topological surface states on mirror-symmetric surfaces due to the nontrivial mirror Chern number, as we see in Appendix \ref{ap:mirrorchern}.

Thus, we conclude that the mass term is a pseudoscalar in HOTIs with $\mu_{1}=2$. 
In contrast, it is straightforward to see that in normal insulators with $\mu_{1}=0$, the mass is a scalar, but not a pseudoscalar. 
Next, we consider cases with combinations between $\mathcal{I}$ symmetry and mirror symmetry.
For example, we consider a system with $\mathcal{I}$ symmetry and $M_{z}$ symmetry with respect to $xy$ plane, leading to the concomitant $C_{2z}$ symmetry.
In this case, the chiral edge mode is along a great circle on the sphere along the mirror plane as shown in  Fig.~\ref{ap_hinge_sphere}(c). It  leads to the gapless surface states on the side surfaces in the rod geometry along the $z$ direction. These gapless surface states are protected by mirror symmetry. Meanwhile if the surfaces of the crystal is not mirror symmetric, CHSs appear.
In the case with $M_{x}$, $M_{y}$ and $M_{z}$ symmetries, the gapless surface states appear in the rod geometries along $x$, $y$ and $z$ directions (Fig.~\ref{ap_hinge_sphere}(d)). These gapless surface states are characterized by the non-trivial mirror Chern number. Here, the mirror Chern number for the $M_{z}$ symmetry is defined as
\begin{equation}\label{appendix_eq_a3}
C_{m}(k_{z})\equiv (C_{+}(k_{z})-C_{-}(k_{z}))/2,
\end{equation}
where $k_{z}$ is $0$ or $\pi$, corresponding to mirror-invariant planes.
In Eq.~\ref{appendix_eq_a3} in spinful systems, $C_{\pm}(k_{z})$ are the Chern numbers in the mirror subspace $M_{z}=\pm i$ defined as
\begin {equation}
C_{\pm}(k_{z})=\frac{1}{2\pi}\int_{\rm BZ}dk_{x}dk_{y}{\rm Tr}[\mathcal{F}^{\pm}_{xy}(\boldsymbol{k})],
\end{equation}
where $\mathcal{F}^{\pm}_{xy}(\boldsymbol{k})$ is the non-Abelian Berry curvature in the $\pm i$ mirror subspace.
In spinless systems, the mirror subspaces are defined by $M_{z}=\pm 1$ instead.
As shown in Appendix~\ref{ap:mirrorchern}, the sum of the mirror Chern numbers in both $k_{z}=0$ and $k_{z}=\pi$ sectors, is a non-zero odd number when $\mu_{1}=2$. Thus, the system with mirror symmetry has gapless surface modes protected by mirror symmetry when $\mu_{1}=2$.

\subsection{Point groups}\label{appendix_pointgroup}
As discussed in the above, the mass term on a sphere includes various information of positions of CHSs. 
In this subsection, we consider the distributions of the mass term on spheres with various symmetries and classify them in terms of  point groups.
As discussed above, the CHS forms a loop protected by $\mathcal{I}$ symmetry in the point group $\overline{1}$  without other symmetries (Type 1, Fig.~\ref{mass}(a-1)).
Here, we add rotation $C_{nz}$ ($n=2,3,4,6$) symmetries around $z$ axis. 
The mass signs do not flip under rotations. In the case with $C_{3z}$ symmetry, that is, in the point group $\overline{3}$, the CHS on the sphere is  invariant under $C_{3z}$ symmetry and $\mathcal{I}$ symmetry as shown in Fig.~\ref{mass}(b) (Type 2).
On the other hand, in the case with $C_{2z}$ symmetry, the combination of $\mathcal{I}$ symmetry and $C_{2z}$ symmetry leads to the mirror $M_{z}$ symmetry with respect to $xy$ plane, and the gapless line is along the great circle on the mirror plane (Fig.~\ref{mass}(c)). The same applies to the case with $C_{4z}$ and $C_{6z}$ symmetries and we call this case Type 3.

In the point groups $mmm$ and $m\overline{3}$, CHSs appear along the three great circles because of $M_{x}$, $M_{y}$ and $M_{z}$ mirror symmetries with respect to $yz$, $zx$ and $xy$ planes, respectively (Type 4, see Fig.~\ref{mass}(d)). In addition, CHSs appear along three great circles which are related by $C_{3z}$ symmetry in the point group $\overline{3}m$ as shown in Fig.~\ref{mass}(e) (Type 5). The point group $4/mmm$ has two mirror symmetries $M_{xy}$ and $M_{x\overline{y}}$ that leave the $x=y$ and the $x=-y$ planes invariant, respectively, in addition to the  mirror symmetries $M_{x}$, $M_{y}$ and $M_{z}$. Because of these mirror symmetries, CHSs appear along the five great circles as shown in Fig.~\ref{mass}(f) (Type 6). In the point groups $6/mmm$ and $m\overline{3}m$, seven and nine great circles of gapless lines appear,  respectively, because of additional mirror and rotation symmetries (Types 7 and 8, shown in Figs.~\ref{mass}(g) and \ref{mass}(h)).

\subsection{Screw symmetry}

Next, we consider the cases with screw symmetry. In this case, the combination between $\mathcal{I}$ symmetry and screw rotation symmetry leads to the mirror symmetry whose mirror plane is offset from the inversion center. For example, we consider the two symmetry operations, $\{\mathcal{I}|000\}$ and $\{C_{2z}|00\frac{1}{2}\}$ corresponding to the type-I MSG No.~11. These symmetries lead to $\{M_{z}|00\frac{1}{2}\}$, whose mirror plane does not include the inversion center.
In this case, when $\mu_{1}=2$ and the Chern number on $k_{z}=0$ plane is zero, the mirror Chern number is nontrivial as discussed in Appendix~\ref{sec:appnedixE:connectionbetweenmirror}. It leads to the emergence of the surface states protected by mirror symmetry, and therefore the mass distribution in this case corresponds to Type 3 in Fig.~\ref{mass}.

\subsection{Glide symmetry}\label{ap_subsec_glide}

In this subsection, we consider the cases with glide symmetry. 
The existence of glide operation invalidates our assumption that the surface orientation specified by $\boldsymbol{n}$ uniquely determines the sign of the mass. For example, let us consider the glide $\{M_{y}|00\frac{1}{2} \}$. Then, if the mass on the $z=0$ surface is positive, the mass on the $z=1/2$ surface is negative. Thus, the mass sign depends on a choice of surface terminations. 
Therefore, we cannot determine the sign of the mass on the $z=0$ surface uniquely by glide symmetry. 

Next, we consider other surface orientations with such ambiguity of the mass sign. Let $(\alpha \beta \gamma)$ denote the Miller index for the surface with this mass ambiguity. It follows that $\beta=0$. The $(\alpha 0\gamma)$ plane can be written as
\begin{equation}\label{glideinvariantplane1}
\alpha x+\beta y +\gamma z =d,
\end{equation}
where $d$ is a constant. This plane is transformed into 
\begin{equation}\label{glideinvariantplane2}
\alpha x-\beta y +\gamma \left(z-\frac{1}{2} \right) =d,
\end{equation}
under the glide $\{M_{y}|00\frac{1}{2} \}$. The mass ambiguity appears when the two surfaces expressed by  Eq.~(\ref{glideinvariantplane1}) and by Eq.~(\ref{glideinvariantplane2}) are parallel but are not equivalent under lattice translation symmetry. Thus, the following equation holds: $\gamma \equiv 1$ (mod 2) and $\beta =0$.
Thus, the $(\alpha, 0,  \gamma)$ surfaces have such ambiguity if $\gamma$ is an odd number, and therefore there is an infinite number of surface orientations on which the mass ambiguity arises.

As discussed in the main text, when $\gamma \equiv 0$ (mod 2) and $\alpha \equiv 1$ (mod 2), the $(\alpha 0 \gamma)$ plane is invariant under the glide $\{M_{y}|00\frac{1}{2} \}$, and therefore the mass on this plane is zero. 
From this, within the glide-mirror plane, two cases for the behavior of the surface mass coexist: 
the mass on the $(\alpha, 0, \gamma)$ surface is zero when $\alpha \equiv 1$ (mod 2) and $\gamma \equiv 0$ (mod 2), while the mass is indeterminate when $\gamma \equiv 1$ (mod 2). Thus, for the surface normal within the glide-mirror plane, the dependence of the mass on the surface normal vector $\bs{n}$ is singular. To specify the singular behavior for $\bs{n}$ perpendicular to the glide plane, we draw broken circles, and we call the cases where the mirror planes in Types 3-8 are replaced with glide mirror as Types $3'$-$8'$ (Figs.~\ref{105glidemass}(b-g)).

\section{Proof of $C^{z}_{\rm slab}=0$ when $N$ is an even number}\label{appnedixb:chernzero}

\begin{figure}[t]
\centerline{\includegraphics[width=1\columnwidth]{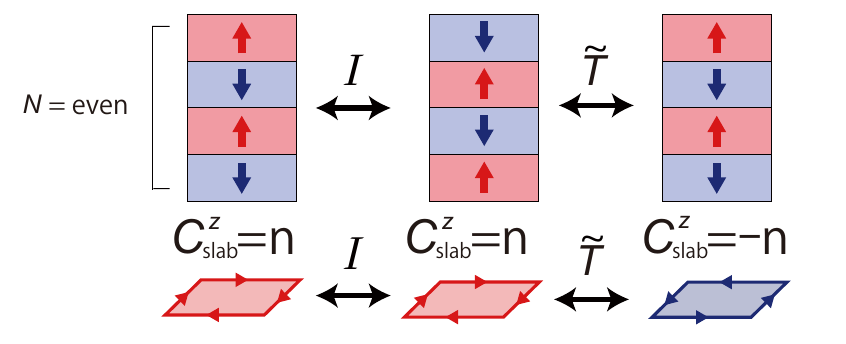}}
  \caption{Two symmetry operations on a slab of a layered AFM with an even number of layers. The slab with even $N$ and its inversion partner connected by the inversion operation $\mathcal{I}$. The Chern number of the two slabs are the same. In addition, the slab with even $N$ and its inversion partner connected by the anti-unitary operation $\tilde{\mathcal{T}}$. The Chern number of the two slabs are of the opposite signs. }
\label{even_Chern}
\end{figure}

Here, we show that a Chern number is $C^{z}_{\rm{slab}}=0$ for a slab of layered AFMs with $\mathcal{I}$ symmetry and $\tilde{\mathcal{T}}=C_{2z} \mathcal{T}$ symmetry with the even number of layers. In the following, $N$ represents a number of the  layers. 
In the following, we show the relation $C^{z}_{\rm slab}=-C^{z}_{\rm slab}$ via the combination of $\mathcal{I}$ and an anti-unitary operation $\tilde{\mathcal{T}}= C_{2z}\mathcal{T}$, where $\mathcal{T}$ represents a time-reversal operation and $C_{2z}$ represents a two-fold rotation around $z$-axis as shown in Fig.~\ref{even_Chern}. Slabs with even $N$ are symmetric under the combination of a unitary operation $\mathcal{I}$ and an anti-unitary operation $\tilde{\mathcal{T}}$: 
\begin{equation}
\tilde{\mathcal{T}} \mathcal{I} \mathcal{H} (\boldsymbol{k}) (\tilde{\mathcal{T}}\mathcal{I})^{-1}= \mathcal{H}(-\boldsymbol{k}),
\end{equation}
where $\mathcal{H}(\boldsymbol{k})$ is the Hamiltonian of the slab with even $N$.
From this, for an occupied state $\ket{u_{n}(\boldsymbol{k})}$ of the Hamiltonian $\mathcal{H}(\boldsymbol{k})$ with an eigenvalues $E_{n}$, the following relation holds:
\begin{align}
&\mathcal{H}(-\boldsymbol{k})\tilde{\mathcal{T}} \mathcal{I} \ket{u_{n}(\boldsymbol{k})}
\nonumber \\
=&\tilde{\mathcal{T}} \mathcal{I}\mathcal{H}(\boldsymbol{k})\ket{u_{n}(\boldsymbol{k})} 
=E_{n}\tilde{\mathcal{T}}\mathcal{I} \ket{u_{n}(\boldsymbol{k})}.
\end{align}
Therefore, $\tilde{\mathcal{T}} \mathcal{I} \ket{u_{n}(\boldsymbol{k})}$ is an eigenstate of the Hamiltonian $\mathcal{H}(-\boldsymbol{k})$, and we can expand 
\begin{equation}
\tilde{\mathcal{T}}\mathcal{I} \ket{u_{n}(\boldsymbol{k})}=\sum_{m}U_{mn}(\boldsymbol{k})\ket{u_{m}(-\boldsymbol{k})}, 
\end{equation}
where $U_{mn}(\boldsymbol{k})$ are the matrix elements of a unitary transformation acting on the space of occupied states.
Because of the anti-unitarity of $\tilde{\mathcal{T}}$, we obtain
\begin{equation}
\ket{u_{n}(\boldsymbol{k})}= \sum_{m} U^{*}_{mn}(\boldsymbol{k})  \ket{\mathcal{I} \tilde{\mathcal{T}} u_{m}(-\boldsymbol{k})}.
\end{equation}

Therefore, the Berry connection is expressed as
\begin{align}
&(\mathcal{A}_{\mu}(\boldsymbol{k}))_{nn'}\nonumber \\
=&i \sum_{m,m'}U_{mn} (\boldsymbol{k}) \bra{\mathcal{I} \tilde{\mathcal{T}} u_{m}(-\boldsymbol{k})}
\partial_{\mu}
\biggl(
U^{*}_{m'n'}(\boldsymbol{k}) \ket{\mathcal{I} \tilde{\mathcal{T}} u_{m'}(-\boldsymbol{k})}\biggr) \nonumber \\
=&- \sum_{m,m'}U^{\dagger}_{n'm'}(\boldsymbol{k}) (\mathcal{A}_{\mu}(-\boldsymbol{k}))_{m'm}U_{mn}(\boldsymbol{k})\nonumber \\
&\ \ \ \ \ \ \ \ \ \ \ \ \ \ \ \ \ \ \ \ \ \ \ \ \ \ \ \ -i\sum_{m}U^{\dagger}_{n'm}(\boldsymbol{k})\partial_{\mu} U_{mn}(\boldsymbol{k})\nonumber \\
=&-(U^{\dagger}(\boldsymbol{k}) (\mathcal{A}_{\mu}(-\boldsymbol{k}))U(\boldsymbol{k}))^{T}_{nn'}\nonumber \\
&\ \ \ \ \ \ \ \ \ \ \ \ \ \ \ \ \ \ \ \ \ \ \ \ \ \ \ \ -i(U^{\dagger}(\boldsymbol{k})\partial_{\mu} U(\boldsymbol{k}))^{T}_{nn'},
\end{align}
and then the Berry curvature is written by 
\begin{equation}
\mathcal{F}_{ab}(\boldsymbol{k})=-(U^{\dagger}(\boldsymbol{k}) \mathcal{F}_{ab}(-\boldsymbol{k}) U(\boldsymbol{k}))^{T}.
\end{equation}

Therefore, we obtain the relationship $C^{z}_{\rm slab}=-C^{z}_{\rm slab}$.
We note that this proof is based on the existence of the anti-unitary operation $\tilde{\mathcal{T}}$, which interchanges the slab with even $N$ and its inversion partner (see Fig.~\ref{even_Chern}). In other layered AFMs with an anti-unitary operator $\tilde{\mathcal{T}}$ such as $\tilde{\mathcal{T}}= C_{3z} \mathcal{T}$, $C_{4z} \mathcal{T}$ or $C_{6z} \mathcal{T}$, the Chern number of the even slab is zero.

\section{Chern number for a 2D slab from a 3D HOTI}\label{appendixc:chernnumber}

According to the previous works \cite{PhysRevB.97.205136, PhysRevB.98.205129, PhysRevResearch.2.013300, wieder2018axioninsulatorpump}, a two-dimensional (2D) slab of a 3D HOTI with inversion symmetry, with a finite thickness along the $z$ direction is a 2D Chern insulator with the Chern number $C^{z}_{\rm slab}=1$ (mod 2). 
In particular, in the case of the layered insulating AFM, when $N={\rm odd}$, $\mathcal{I}$ symmetry is preserved, and the slab Chern number $C^{z}_{\rm slab}$ (mod 2) satisfies \cite{PhysRevResearch.2.013300}; 
\begin{equation}
C^{z}_{\rm slab}\equiv \frac{1}{2}\mu_{1}\ \ ({\rm mod}\ 2),
\end{equation}
where $\mu_{1}=0$ or 2 (mod 4). Therefore, when $\mu_{1}=2$ in the bulk, the slab system with odd $N$ is a 2D Chern insulator with $C^{z}_{\rm slab}=1$ (mod 2), $i.e.$ $C^{z}_{\rm slab}=2M+1$.

\begin{figure}
\centerline{\includegraphics[width=1\columnwidth]{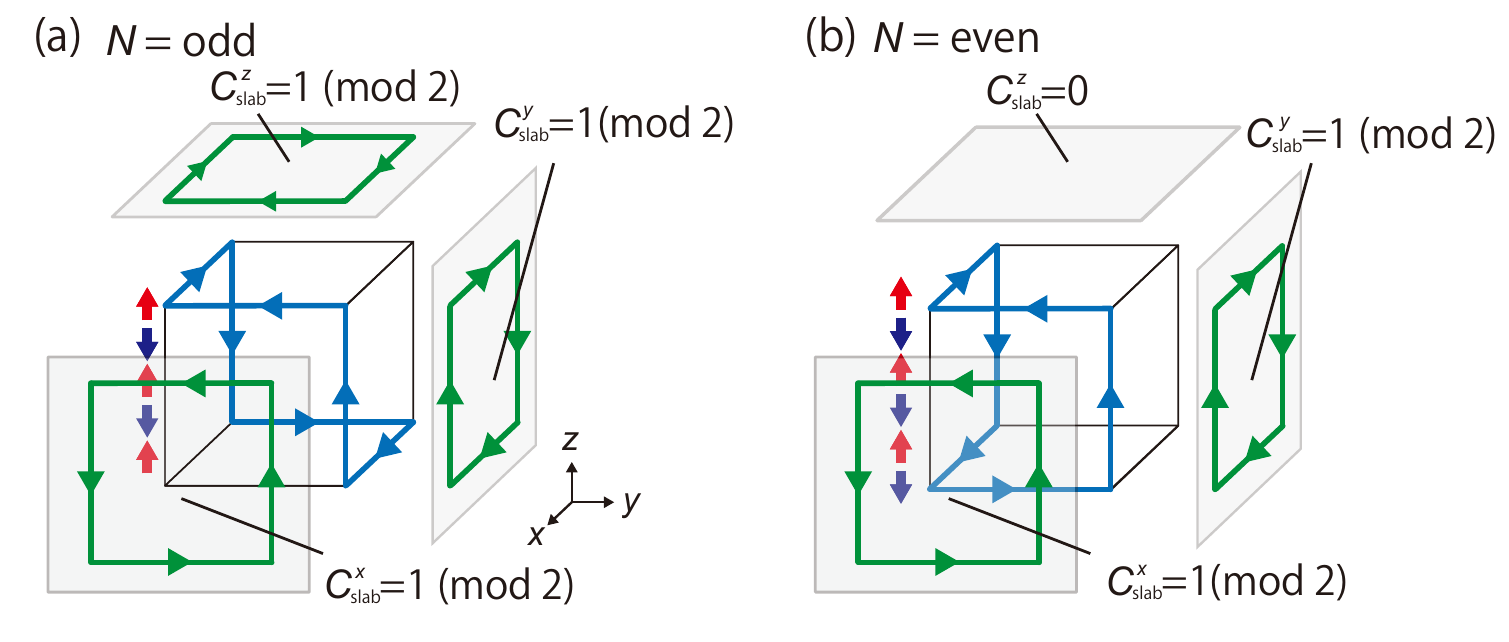}}
  \caption{The Chern numbers of the two slabs with odd and even $N$.  (a) Chern numbers in the slabs with a finite system size along the $x$, $y$ and $z$ directions are $C^{x}_{\rm slab}=C^{y}_{\rm slab}=C^{z}_{\rm slab}=1$ in the case with an odd number of the layers. (b) In the case with even number of the layers, $C^{x}_{\rm slab}=C^{y}_{\rm slab}=1$ and $C^{z}_{\rm slab}=0$, which leads to topological hinge states, $i.e.$ IAHS.}
\label{projection}
\end{figure}

In such a case, we can choose $C^{z}_{\rm slab}=1$ without losing generality. 
One can simultaneously attach two 2D Chern insulators with the same Chern number on two surfaces of the opposite sides of the crystal, so that $\mathcal{I}$ symmetry is preserved\cite{schindler2018higher,PhysRevB.97.205136,PhysRevB.98.205129}.
From this, a Chern number = $2M+1$ can be transformed into a Chern number = $1$ while preserving $\mathcal{I}$ symmetry. Thus, in the main text, we choose the $C^{z}_{\rm slab}=1$.

In addition, we consider Chern numbers in the slabs with a finite system size along the $x$ and $y$ directions; $C^{x}_{\rm slab}$ and $C^{y}_{\rm slab}$. 
Here we assume the crystal shapes are $\mathcal{I}$ symmetric. 
In this case, we obtain the Chern numbers $C^{x}_{\rm slab}$ and $C^{y}_{\rm slab}$ form the topological $\mathbb{Z}_{4}$ index $\mu_{1}$ as
\begin{equation}
C^{x}_{\rm slab}\equiv \frac{1}{2}\mu_{1}\ \ \ C^{y}_{\rm slab}\equiv \frac{1}{2}\mu_{1}\ \ \ ({\rm mod}\ 2),
\end{equation}
where $\mu_{1}$ is even \cite{PhysRevB.98.205129, PhysRevResearch.2.013300}. Therefore, the HOTI with $\mu_{1}=2$ is a 2D Chern insulator in the case with a finite thickness along the $x$ and $y$ directions. Indeed, as shown in Fig.~\ref{projection}(a) and \ref{projection}(b), the projections of CHSs onto the $yz$ or $xz$ planes form loops corresponding to chiral edge states due to the Chern numbers
$
C^{x}_{\rm slab}\equiv C^{y}_{\rm slab}\equiv 1 \ \ ({\rm mod}\ 2),
$
in both cases with odd and even $N$. Therefore, IAHS (for even $N$) are topological gapless states characterized by 
\begin{equation}
C^{x}_{\rm slab}\equiv C^{y}_{\rm slab}\equiv 1\ {\rm and}\  C^{z}_{\rm slab}\equiv 0\ \ ({\rm mod}\ 2).
\end{equation}
This topological properties of IAHS are different from that of conventional $\mathcal{I}$-symmetric hinge states for odd $N$, $i.e.$
$
C^{x}_{\rm slab}\equiv C^{y}_{\rm slab}\equiv C^{z}_{\rm slab}\equiv 1.
$
We note that the relationship $C^{x}_{\rm slab}\equiv C^{y}_{\rm slab}\equiv 1$ (mod 2) is due to the bulk topology $\mu_{1}=2$, and the 2D Chern numbers remain nontrivial as long as the topological invariant $\mu_{1}$ in the bulk is preserved.

\section{Connection between mirror Chern numbers and $\mu_{1}$}\label{ap:mirrorchern}
In this appendix, we discuss connection between the mirror Chern numbers and the symmetry-based indicator $\mu_{1}$. 
Here, we consider a 3D magnetic insulator with $\mathcal{I}$ symmetry and $M_{z}$ mirror symmetry.
First, we define the mirror Chern numbers in the $k_{z}=0$ and $k_{z}=\pi$ sectors as 
\begin{equation}\label{donemirrorcherndefine}
C^{0}_{m}\equiv \frac{1}{2}\bigl( C^{0}_{+}-C^{0}_{-}\bigr),\ \ C^{\pi}_{m}\equiv \frac{1}{2}\bigl( C^{\pi}_{+}-C^{\pi}_{-}\bigr)
\end{equation}
respectively, where $C^{0}_{\pm}$ and $C^{\pi}_{\pm}$ represent the Chern numbers in the mirror sectors with mirror eigenvalues $\pm i$ in the $k_{z}=0$ and $k_{z}=\pi$ sectors respectively.
In the following, we prove that the $\mathbb{Z}_{4}$ symmetry-based indicator $\mu_{1}$ in an insulator is related with mirror Chern number via
\begin{equation}\label{eq:mirrorchern_and_mu1}
\mu_{1}\equiv 2(C^{0}_{m}+C^{\pi}_{m})\ \  ({\rm mod\ 4}),
\end{equation}
where $\mu_{1}=0,2$. The values $\mu_{1}=0,2$ means that the bulk is insulating, while $\mu_{1}=1,3$ corresponds to the Weyl semimetal phase. Equation~(\ref{eq:mirrorchern_and_mu1}) means that when  $\mu_{1}=2$, one of the two mirror Chern numbers $C^{0}_{m}$ and $C^{\pi}_{m}$, is an odd number, and the other is an even number, which means emergence of topological surface states on a mirror-symmetric surface.

In the following, we prove Eq.~(\ref{eq:mirrorchern_and_mu1}).
First, let $C_{m}$ denote the quantity of the r.h.s.  of Eq.~(\ref{eq:mirrorchern_and_mu1});
\begin{equation}
C_{m} \equiv C^{0}_{m}+C^{\pi}_{m}.
\end{equation}
Next, we get the relation
\begin{equation}\label{cherninsulator}
C^{0}_{+}+C^{0}_{-}=C^{\pi}_{+}+C^{\pi}_{-},
\end{equation}
because in an insulator the total Chern number on the $k_{z}={\rm const}$ plane is independent of the values of $k_{z}$. In addition, from Ref.~[\onlinecite{PhysRevB.86.115112}], the following two equations hold:
\begin{equation}\label{paritychern1}
{\rm exp}[i\pi C^{0}_{+}]=\prod_{\Lambda_{i}}\xi^{(+)}(\Lambda_{i},0),
\end{equation}
\begin{equation}\label{paritychern2}
{\rm exp}[-i\pi C^{\pi}_{-}]=\prod_{\Lambda_{i}}\xi^{(-)}(\Lambda_{i},\pi),
\end{equation}
where $\xi^{(\pm)}(\Lambda_{i},k_{z})$ is the parity eigenvalues of the occupied states in $\pm$ mirror sectors, and $\Lambda_{i}=(k_{x},k_{y})$ represents 2D TRIM. From Eqs.~(\ref{cherninsulator}), (\ref{paritychern1}) and (\ref{paritychern2}), the following equation holds:
\begin{align}
{\rm exp}[ i\pi C_{m}]&={\rm exp}[ i\pi (C^{0}_{+}-C^{\pi}_{-})]\nonumber \\
&=\prod_{\Lambda_{i}} \xi^{(+)}(\Lambda_{i},0)\xi^{(-)}(\Lambda_{i},\pi).
\end{align}
Next, let $N^{(\beta)}_{\alpha}$ ($\alpha=\pm$, $\beta=\pm$) the number of irreducible representations (irreps) with $\alpha$ parity in the $\beta$ mirror sector. Then we obtain
\begin{equation}\label{nu/2}
C_{m} \equiv \sum_{\Lambda_{i}}\Bigl( N^{(+)}_{-}(\Lambda_{i},0)+N^{(-)}_{-}(\Lambda_{i}, \pi) \Bigr)\ {(\rm mod\ 2)}.
\end{equation}

By the way, the $M_{z}$ mirror eigenvalue is common within $k_{z}=0$ plane and it is also the case within $k_{z}=\pi$ plane. Therefore the number of the irreps in the $(+)$ mirror sector, $N^{(+)}_{+} (\Lambda_{i},0)+N^{(+)}_{-} (\Lambda_{i}, 0)$ is independent of $\Lambda_{i}$. It follows that
\begin{equation}\label{commonmirrorzero}
\sum_{\Lambda_{i}}\Bigl( N^{(+)}_{+} (\Lambda_{i},0)+N^{(+)}_{-} (\Lambda_{i}, 0) \Bigr)\equiv 0\ ({\rm mod\ 4}).
\end{equation} 
Similarly, in the $(-)$ mirror sector, the following equation holds:
\begin{equation}\label{commonmirrorpi}
\sum_{\Lambda_{i}}\Bigl( N^{(-)}_{+} (\Lambda_{i},\pi)+N^{(-)}_{-} (\Lambda_{i}, \pi) \Bigr)\equiv 0\ ({\rm mod\ 4}).
\end{equation}
By using the Eqs.~(\ref{nu/2}), (\ref{commonmirrorzero}) and (\ref{commonmirrorpi}), we get the following equation:
\begin{align}\label{nu/2ver2}
C_{m}\equiv &\frac{1}{2}\sum_{\Lambda_{i}}\Bigl( N^{(+)}_{-}(\Lambda_{i},0) - N^{(+)}_{+}(\Lambda_{i},0)\nonumber \\
&\ \ \ +N^{(-)}_{-}(\Lambda_{i}, \pi)-N^{(-)}_{+}(\Lambda_{i},\pi) \Bigl)\ ({\rm mod\ 2}).
\end{align}
Then, the $C_{2z}$ eigenvalue is common between $\boldsymbol{k}=(\Lambda_{i},0)$ and $\boldsymbol{k}=(\Lambda_{i}, \pi)$. Therefore, the total number of irreps with the positive eigenvalue $+i$ ($+1$) of $C_{2z}$ at $(\Lambda_{i},0)$ for spinful fermions (spinless fermions), is the same as the number with the same eigenvalue at $(\Lambda_{i}, \pi)$. By noting ${C}_{2z}= {M}_{z} \mathcal{I}$, we obtain 
\begin{align}\label{c2zeigenvaluesinvariant}
N^{(+)}_{+}(\Lambda_{i},0)+N^{(-)}_{-}(\Lambda_{i},0)\nonumber \\
=N^{(+)}_{+}(\Lambda_{i},\pi)+N^{(-)}_{-}(\Lambda_{i},\pi).
\end{align}
The same is true for the number of irreps with the negative eigenvalue of $C_{2z}$:
\begin{align}\label{c2zeigenvaluesinvariantver2}
N^{(+)}_{-}(\Lambda_{i},0)+N^{(-)}_{+}(\Lambda_{i},0)\nonumber \\
=N^{(+)}_{-}(\Lambda_{i},\pi)+N^{(-)}_{+}(\Lambda_{i},\pi).
\end{align}
From Eqs.~(\ref{nu/2ver2}), (\ref{c2zeigenvaluesinvariant}) and (\ref{c2zeigenvaluesinvariantver2}), the following equation holds:
\begin{align}
C_{m}\equiv &\frac{1}{4}\sum_{\Lambda_{i}:{\rm 2D\ TRIM}}\Bigl[ N^{(+)}_{-}(\Lambda_{i},0)+N^{(-)}_{-}(\Lambda_{i},0)\nonumber \\
&-N^{(+)}_{+}(\Lambda_{i},0)-N^{(-)}_{+}(\Lambda_{i},0) +N^{(+)}_{-}(\Lambda_{i},\pi)\nonumber \\
&+N^{(-)}_{-}(\Lambda_{i},\pi)-N^{(+)}_{+}(\Lambda_{i},\pi)-N^{(-)}_{+}(\Lambda_{i},\pi)\Bigr] \ ({\rm mod\ 2})\nonumber \\
=&\frac{1}{4}\sum_{\Gamma_{i}:{\rm 3D\ TRIM}} \Bigl[N_{-}(\Gamma_{i})-N_{+}(\Gamma_{i}) \Bigr]
\end{align}
Thus we obtain Eq.~(\ref{eq:mirrorchern_and_mu1}).

\section{Connection between mirror Chern numbers and $\mu_{1}$ in the presence of screw symmetry}\label{sec:appnedixE:connectionbetweenmirror}

In this appendix, we discuss connection between the mirror Chern numbers and $\mu_{1}$ in the presence of screw $\overline{C}_{2z} \equiv \{C_{2z}|00\frac{1}{2}\}$ symmetry. Here, we consider a 3D magnetic insulator with $\mathcal{I}$ symmetry and screw $\overline{C}_{2z}$ symmetry. In addition, the combination between the $\mathcal{I}$ symmetry and the screw symmetry leads to a mirror symmetry $\overline{M}_{z}\equiv \{M_{z}|00\frac{1}{2} \}$, where mirror plane $z=\frac{1}{4}$ does not include the inversion center at $(0, 0, 0)$. In the following, we prove that the mirror Chern number (see Eq.~(\ref{donemirrorcherndefine})) with respect to this mirror symmetry is related with the symmetry-based indicator $\mu_{1}$ via
\begin{equation}\label{screwmirrorchernnumbertomuone}
\mu_{1}\equiv 2(C^{0}_{m}+C^{\pi}_{m})=2C^{0}_{m}\ \ ({\rm mod \ 4}),
\end{equation}
when the Chern number on $k_{z}=0$ plane is zero. Equation (\ref{screwmirrorchernnumbertomuone}) means that when $\mu_{1}=2$, topological surface states on a mirror-symmetric surface appear. 

In the following, we prove Eq.~(\ref{screwmirrorchernnumbertomuone}). First, we consider the plane $k_{z}=\pi$. In this case, $\overline{C}_{2z}$ operation and $\overline{M}_{z}$ operation anticommute:
\begin{equation}
\overline{C}_{2z} \overline{M}_{z} =- \overline{M}_{z}\overline{C}_{2z}.
\end{equation}
Therefore, an occupied state $\ket{u_{n}(\boldsymbol{k})}$ and another occupied state $\overline{C}_{2z} \ket{u_{n}(\boldsymbol{k})}$ have the opposite mirror eigenvalues to each other. Here, the number $\nu$ of occupied states is an even number.  
Then the Chern number of $\ket{u_{n}(\boldsymbol{k})}$ is equal to that of $\overline{C}_{2z} \ket{u_{n}(\boldsymbol{k})}$, and these two states are in the opposite mirror sectors. 
Therefore, we get the equation $C^{\pi}_{+} = C^{\pi}_{-}$. In other words, the mirror Chern number in the plane $k_{z}=\pi$ is zero:
\begin{equation}\label{eqappendiEmirrorchernzeropimen}
C^{\pi}_{m}=0.
\end{equation} 
Similarly, $\mathcal{I}$ operation and $\overline{M}_{z}$ operation anticommute, and then an occupied state $\ket{u_{n}(\boldsymbol{k})}$ and another state $\overline{M}_{z} \ket{u_{n}(\boldsymbol{k})}$ have the opposite parity eigenvalues each other, and they are degenerate. Therefore, the number of the irreps with even parity $N_{+} (\Lambda_{i}, \pi)$ is the same as that with odd parity $N_{-} (\Lambda_{i}, \pi)$:
\begin{equation}\label{apescreweparitysame}
\sum_{\Lambda_{i}}\ N_{+}(\Lambda_{i},\pi)=\sum_{\Lambda_{i}}N_{-}(\Lambda_{i},\pi),
\end{equation}
where  $\Lambda_{i}=(k_{x}, k_{y})$ runs through four 2D TRIM.

Next, we consider the plane $k_{z}=0$. Unlike $k_{z}=\pi$ plane, $\mathcal{I}$ operation and $\overline{M}_{z}$ operation commute on $k_{z}=0$ plane. 
Let $N^{(\beta)}_{\alpha}$ ($\alpha=\pm$, $\beta=\pm$) the number of irreps with $\alpha$ parity in the $\beta$ mirror sector.
Because the $M_{z}$ mirror eigenvalue is common within $k_{z}=0$ plane,  the following equation holds:
\begin{equation}\label{appendixEcommonmirrorplus}
\sum_{\Lambda_{i}}\Bigl( N^{(+)}_{+} (\Lambda_{i},0)+N^{(+)}_{-} (\Lambda_{i}, 0) \Bigr)\equiv 0\ ({\rm mod\ 4}).
\end{equation} 
In addition, on the line $(k_{x},k_{y})=\Lambda_{i}$, the  $\overline{C}_{2z}$ symmetry is preserved, and the eigenstates are classified into two $\overline{C}_{2z}$ sectors with $\overline{C}_{2z}=\pm e^{-i k_{z}/2}$.  At the 3D TRIM $(\Lambda_{i}, \pi)$, $\overline{C}_{2z}$ and $\mathcal{I}$ anticommute, and therefore the number of occupied states with $\overline{C}_{2z}=\pm e^{-ik_{z}/2}$ is $\nu /2$. Thus we conclude that at the 3D TRIM $(\Lambda_{i},0)$, the number of occupied states with $\overline{C}_{2z}=\pm e^{-ik_{z}/2}$ is $\nu /2$. By noting $\overline{C}_{2z}= \overline{M}_{z} \mathcal{I}$, we obtain 
\begin{equation}\label{screwe4}
\sum_{\Lambda_{i}}\left( N^{(+)}_{+}(\Lambda_{i},0)+N^{(-)}_{-}(\Lambda_{i},0)\right)=2\nu,
\end{equation}
\begin{equation}\label{screwe5}
\sum_{\Lambda_{i}}\left( N^{(+)}_{-}(\Lambda_{i},0)+N^{(-)}_{+}(\Lambda_{i},0)\right)=2\nu.
\end{equation}
From Eqs.~(\ref{apescreweparitysame}), (\ref{screwe4}) and (\ref{screwe5}), the following equation holds:
\begin{align}\label{appendixenoe8}
\mu_{1}\equiv \frac{1}{2}\sum_{\Lambda_{i}}&\Bigl[ N^{(+)}_{+}(\Lambda_{i},0) +  N^{(-)}_{+}(\Lambda_{i},0)\nonumber \\
&-N^{(+)}_{-}(\Lambda_{i},0) -  N^{(-)}_{-}(\Lambda_{i},0) \Bigr]\ \ ({\rm mod}\ 4)\nonumber \\
= \sum_{\Lambda_{i}}&\Bigl[ N^{(+)}_{+}(\Lambda_{i},0) -N^{(+)}_{-}(\Lambda_{i},0) \Bigr].
\end{align}
In addition, from Eqs.~(\ref{appendixEcommonmirrorplus}) and (\ref{appendixenoe8}), we get the following equation:
\begin{align}\label{appendixE:equatione9cmu}
\frac{\mu_{1}}{2}&\equiv-\sum_{\Lambda_{i}}N^{(+)}_{-}(\Lambda_{i}, 0)\ \ ({\rm mod}\ 2)\nonumber \\
&\equiv C_{+}^{0}\ \ ({\rm mod}\ 2).
\end{align}

Here we note that the Chern number along the $xy$ plane is even because $C=C_{+}^{\pi}+C^{\pi}_{-}\equiv \sum_{\Lambda_{i}}N_{-}(\Lambda_{i}, \pi) =\frac{\nu}{2}\times 4=2 \nu\equiv 0$ (mod 2) from Eq.~(\ref{apescreweparitysame}). Therefore, from Eq.~(\ref{appendixE:equatione9cmu}), we get 
\begin{align}
\frac{\mu_{1}}{2}\equiv C_{+}^{0}&=\frac{1}{2}(C^{0}_{+}+C^{0}_{-})+\frac{1}{2}(C^{0}_{+}-C^{0}_{-}) \nonumber \\
&=\frac{1}{2}C+C^{0}_{m}\ \ ({\rm mod}\ 2),
\end{align}
which relates the three topological invariants $\mu_{1}$, $C$ and $C^{0}_{m}$. In particular, when we assume that the Chern number on $k_{z}=0$ plane is $C\equiv C^{0}_{+}+C^{0}_{-}=0$ as we adopted in our classification of the CHSs, it follows that
\begin{equation}
\frac{\mu_{1}}{2}\equiv C^{0}_{m}\ \ ({\rm mod}\ 2).
\end{equation}
From this equation and Eq.~(\ref{eqappendiEmirrorchernzeropimen}), we obtain Eq.~(\ref{screwmirrorchernnumbertomuone}).

\end{document}